\documentclass{article}

\usepackage{amsmath}
\usepackage{graphicx}
\usepackage{amsfonts}
\usepackage{amssymb}

\newtheorem{theor}{Theorem}[section]
\newtheorem{cor}{Corrolary}[section]
\newtheorem{pr}{Proposition}[section]
\newtheorem{lm}{Lemma}[section]

\begin{document}

\title{A Lagrangian form of tangent forms}
\author{Paul Popescu}
\date{}
\maketitle

\begin{abstract}
The aim of the paper is to study some dynamic aspects coming from a tangent
form, i.e. a time dependent differential form on a tangent bundle. The
action on curves of a tangent form is natural associated with that of a
second order Lagrangian linear in accelerations, while the converse
association is not unique. An equivalence relation of tangent form,
compatible with gauge equivalent Lagrangians, is considered. We express the
Euler-Lagrange equation of the Lagrangian as a second order Lagrange
derivative of a tangent form, considering controlled and higher order
tangent forms. Hamiltonian forms of the dynamics generated are given,
extending some quantization formulas given by Lukierski, Stichel and
Zakrzewski. Using semi-sprays, local solutions of the E-L equations are
given in some special particular cases.
\end{abstract}



\section{Introduction}

The second order Lagrangians are considered, for example, in \cite{Kra, KP}, 
\cite{NFS}, \cite{To01} etc. (see, for example, \cite{Mi, PMP4, PR} for a
study of higher order Lagrangians). The second order Lagrangians that are
affine in acceleration are involved in some special problems and studied for
example in \cite{Ac}, \cite{CMR}, \cite{CMR02}, \cite{Kra}, \cite{KP}, \cite%
{LSZ01}, \cite{Ma02}, \cite{Ma03}, \cite{PM01} etc. These are the most
possible singular Lagrangians - their vertical hessian vanishes. According
to \cite[Sect. 6.3]{Kra}, some special regularity conditions can be
considered. Third order Lagrangians, that are affine in the third order
derivatives and possessing an acceleration-extended Galilean symmetry, are
studied in \cite{LSZ02}; they extend the second order case considered
previously by the same authors and considered in a general form in this
paper. It can be a model for a future development of constructions in the
present paper.

The goal of this paper is to study tangent forms, i.e. differentiable one
forms $\omega $ on $I\!\!R\times TM$, where $M$ is a manifold. Some basic
aspects and motivating examples can be found in our previous paper \cite%
{PMP5}. We consider an action of a tangent form $\omega $ on differentiable
curves on $M$, in fact the same as the action of a suitable second order
Lagrangian affine in accelerations (it corresponds canonically to $\omega $,
by Proposition \ref{pr P-L}). Conversely, the action of a second order
Lagrangian affine in accelerations can correspond to at least one tangent
form (Proposition \ref{pr L-P}). We consider a certain equivalence relation
on tangent forms such that an equivalence class corresponds to some gauge
equivalent Lagrangians given by the actions (Proposition \ref{preqg}).

Considering controlled tangent forms (Proposition \ref{prdefconder}), higher
order tangent forms, top tangent forms and Lagrange derivatives of tangent
forms, then the Euler Lagrange equation of a tangent form can be obtained by
(two) successive Lagrange derivatives of tangent forms (Proposition \ref%
{pr-ostr}), considering also an Ostrogradski tangent form, closed related to
Ostrogradski momenta. The Euler-Lagrange equation contains the second
derivatives and we prove that in the case of a regular Lagrangian, the
solutions are integral curves of a global second order differential equation
(Proposition \ref{PrPf-regular}).

Considering a Legendre map and considering non-degenerated,
hyper-non-dege\-ne\-rated and biregular Lagrangians, we study the dynamics
given by Lagrangians affine in accelerations given by tangent forms. We
prove that for a regular tangent form, the dynamics on $M$ (i.e. the
solutions of E-L equations) comes from the projection of the integral curves
of a vector field $X$ on $T^{2\ast }M=T^{\ast }M\times _{M}TM$ (Proposition %
\ref{PrPf-non-deg}), while for a biregular tangent form, the dynamics on $M$
comes from the projection of the integral curves of a vector field $Y$ on $%
T_{2}^{0}M=TM\times _{M}TM$ (Proposition \ref{PrPf-non-deg1}).

Important tools in describing the dynamic equations of a Hamiltonian system
are offered by quantization. Following similar ideas used in \cite[Section 2.%
]{LSZ01}, where Ostrogradski-Dirac and Fadeev-Jakiw methods are used, we use
here a modified Ostrogradski-Dirac method, offered by the possibility to
construct constraints slight different from the canonical ones used in
Ostrogradski theory. The Ostrogradski-Dirac method was also used in \cite%
{CMR} to a quantization of a system derived from a Lagrangian affine in
accelerations, involved in the study of a Reegge-Teitelboim model. Since in
the cases considered in our paper it is not necessary to express the
constraints techniques explicitly, we use a symplectic formalism instead,
giving here a global form of the quoted methods. In Subsection \ref%
{subsect-Hamilton} we present a Hamiltonian description of the dynamics
defined by the vector fields $X$ and $Y$ described above, proving that:

-- If $\omega $ is regular and its essential part is time independent, then
there are a symplectic form $\Xi ^{\prime }$ on $T^{2\ast }M$ and a
Hamiltonian $H:I\!\!R\times T^{2\ast }M\rightarrow I\!\!R$ such that the
Hamiltonian vector field $X_{H}$ is $X$ (Theorem \ref{thquant01}).

-- If $\omega $ is biregular and its essential part is time independent,
then there are a symplectic form $\Xi ^{\prime \prime }$ on $T_{2}^{0}M$ and
a Hamiltonian $H^{\prime }:I\!\!R\times T_{2}^{0}M\rightarrow I\!\!R$ such
that the Hamiltonian vector field $X_{H^{\prime }}$ is $Y$ (Theorem \ref%
{thquant02}).

Some examples and special cases are given in Subsection \ref{subsectEx}. In
the case when $\dim M=1$, we prove in Proposition \ref{prM=1} that the
generalized Euler-Lagrange equation of a regular and basic tangent form
admits locally standard Lagrangian descriptions (in the sense of \cite[%
Section 2.]{CN}). In order to describe the dynamics generated by some
classes of tangent forms, we use first order semi-sprays. Following some
concrete examples, we consider some special cases (Propositions \ref{prex01}
to \ref{prex05}) when families of local semi-sprays of first order are
considered; their integral curves project on (sometimes all) integral curves
of the generalized Euler-Lagrange equation associated with the Lagrangian of
the tangent form.

Using local calculus, certain geometrical objects on higher order tangent
bundles and on general fibered manifolds are described in an Appendix.

\section{Tangent forms, Lagrangians and actions on curves}

A \emph{tangent form} on a differentiable manifold $M$ is a differentiable
form $\omega \in \mathcal{X}^{\ast }(I\!\!R\times TM)$. Denote by $%
p_{1}:I\!\!R\times TM\rightarrow I\!\!R$ and $p_{2}:I\!\!R\times
TM\rightarrow TM$ the natural projections. The pull-backs $p_{1}^{\ast }$
and $p_{2}^{\ast }$ of $1$--forms in $\mathcal{X}^{\ast }(I\!\!R\times TM)$
give rise to a direct sum decomposition $\mathcal{X}^{\ast }(I\!\!R\times
TM)=\mathcal{X}_{1}^{\ast }(I\!\!R\times TM)\oplus \mathcal{X}_{2}^{\ast
}(I\!\!R\times TM)$; let $\omega =\omega _{0}+\omega ^{\prime }$ the
corresponding decomposition of $\omega $. We say that $\omega _{0}$ is the 
\emph{Lagrangian component} and $\omega ^{\prime }$ is the \emph{essential
component} of $\omega $.

There is a natural flip $\iota :TT^{\ast }M\rightarrow T^{\ast }TM$ that is
a diffeomorphism of manifolds (see the Appendix). The natural projection $%
\pi _{T^{\ast }M}:TT^{\ast }M\rightarrow T^{\ast }M$ gives the morphism of
vector bundles $\pi _{T^{\ast }M}\circ \iota ^{-1}:T^{\ast }TM\rightarrow
T^{\ast }M$ that induces an epimorphism of vector bundles $C_{TM}=\pi
_{T^{\ast }M}\circ \iota ^{-1}:T^{\ast }TM\rightarrow \pi _{TM}^{\ast
}T^{\ast }M$, over $TM$. Let us call $\ker C_{TM}\subset T^{\ast }TM$ as the 
\emph{co-vertical bundle} of $TM$. It is easy to see that it is canonically
isomorphic with the dual of the vertical vector bundle of $TM$. Indeed, $%
\ker C_{TM}$ is canonically isomorphic with the induced vector bundle $\pi
_{TM}^{\ast }T^{\ast }M$ that is the dual vector bundle of $\pi _{TM}^{\ast
}TM$, canonically isomorphic to its turn with the vertical vector bundle $%
VTM $ of $TM$. (Notice that $VTM$ is the kernel of the differential map of $%
\pi _{TM}:TM\rightarrow M$.) Thus we can denote $\ker C_{TM}=V^{\ast }TM$,
without any confusion. Notice also that using the canonical isomorphism
depicted above of $\pi _{TM}^{\ast }T^{\ast }M$ and $\ker C_{TM}$, then $%
C_{TM}$ gives a canonic map $J^{\ast }:T^{\ast }TM\rightarrow V^{\ast
}TM\subset T^{\ast }TM$ having the property that $\ker J^{\ast }=J^{\ast
}(T^{\ast }TM)=V^{\ast }TM$, thus $(J^{\ast })^{2}=0$; it is the dual
counterpart of the almost tangent structure on $TM$ (see \cite{Le}).

We can consider the time dependent counterpart, taking $T^{\ast
}(I\!\!R\times TM)\rightarrow I\!\!R\times TM$ instead of $T^{\ast
}TM\rightarrow TM$ and $I\!\!R\times V^{\ast }TM\subset T^{\ast
}(I\!\!R\times TM)$, with the base $I\!\!R\times TM$, instead of $V^{\ast
}TM\subset T^{\ast }TM$ with the base $TM$. More specifically, taking into
account the canonical isomorphisms depicted above, the vector bundle $%
I\!\!R\times V^{\ast }TM\rightarrow I\!\!R\times TM$ is canonically
isomorphic with $I\!\!R\times \pi _{TM}^{\ast }T^{\ast }M\rightarrow
I\!\!R\times TM$.

A \emph{top tangent form} $\eta $ is defined as a section of the vector
bundle $I\!\!R\times V^{\ast }TM\rightarrow I\!\!R\times TM$, or, via
canonical isomorphisms, a section of the vector bundle $I\!\!R\times \pi
_{TM}^{\ast }T^{\ast }M\rightarrow I\!\!R\times TM$. Thus we can regard $%
\eta :I\!\!R\times TM\rightarrow \pi ^{\ast }T^{\ast }M$. Since a tangent
form is actually a section $\omega :I\!\!R\times TM\rightarrow I\!\!R\times
T^{\ast }TM$, then it gives a top tangent form $\eta =(I_{I\!\!R}\times
J^{\ast })\circ \omega :I\!\!R\times TM\rightarrow I\!\!R\times VTM$.

Using local coordinates (see the Appendix) a tangent form $\omega $ has the
local expression 
\begin{equation}
\omega =\omega _{0}(t,x^{i},y^{i})dt+(\omega _{i}(t,x^{j},y^{j})dx^{i}+\bar{%
\omega}_{i}(t,x^{j},y^{j})dy^{i})=\omega _{0}+\omega ^{\prime }.
\label{eqlocom}
\end{equation}%
A top tangent form $\eta $ has the local expression $\eta =\eta
_{i}(t,x^{j},y^{j})dx^{i}$; the top tangent form given by the tangent form (%
\ref{eqlocom}) has the expression $\bar{\omega}=\bar{\omega}_{i}dx^{i}$.

A tangent form can be related to a second order dynamic form considered in 
\cite{Kra}. According to \cite[Section 2]{Kra}, a \emph{first order dynamic
form} on the bundle $Y=I\!\!R\times M\rightarrow M$ is a one contact and
horizontal two form $\nu $ on $J^{1}(Y)$, having the local expression $\nu
=\nu _{i}(t,x^{j},y^{j})dx^{i}\wedge dt+\bar{\nu}_{i}(t,x^{j},y^{j})dy^{i}%
\wedge dt$. Obviously a first order dynamic form is equivalent to give a
pure tangent form. An advantage to use tangent forms is having the
Lagrangian forms in the same setting. Another motivation to use tangent
forms is given by their action on curves, the same as the action of suitable
second order Lagrangians that are affine in accelerations.

If $\gamma :[a,b]\rightarrow M$ is a curve on $M$, then for $t\in \lbrack
a,b]$, consider $\tilde{\gamma}(t)=(t,\frac{d\gamma }{dt}(t))\in
I\!\!R\times T_{\gamma (t)}M$ and the scalar $\omega _{\tilde{\gamma}%
(t)}\left( \frac{d^{2}\gamma }{dt^{2}}(t)\right) $. The \emph{action} of the
tangent form $\omega $ on $\gamma $ is given by the formula%
\begin{equation}
I_{\omega }(\gamma )=\int_{a}^{b}\omega _{\tilde{\gamma}(t)}\left( \frac{%
d^{2}\gamma }{dt^{2}}(t)\right) dt.  \label{action001}
\end{equation}

Using local coordinates: $t\in I\!\!R$, $(x^{i})$ on $M$ and $(x^{i},y^{i})$
on $TM$, if $\omega $ has the expression (\ref{eqlocom}) and a curve $\gamma 
$ has $t\rightarrow (x^{i}(t))$, then the action (\ref{action001}) has the
expression 
\begin{equation}
I_{\omega }(\gamma )=\int_{a}^{b}(\omega _{0}+\omega _{i}{{\frac{dx^{i}}{dt}}%
}+\bar{\omega}_{i}{{\frac{d^{2}x^{i}}{dt^{2}}}})dt.  \label{action001a}
\end{equation}

Let us relate the action of tangent forms on curves to the actions of
Lagrangians on curves. First, the \emph{action} of $%
L^{(1)}:I\!\!R\times TM\rightarrow I\!\!R$ on a curve $\gamma
:[a,b]\rightarrow M$ is given by the formula:%
\begin{equation*}
I_{L}(\gamma )=\int_{a}^{b}L^{(1)}\left( t,\gamma (t),\frac{d\gamma }{dt}%
(t)\right) dt.
\end{equation*}

If $\gamma :[a,b]\rightarrow M$ is a curve on $M$, then the curves $\frac{%
d\gamma }{dt}:[a,b]\rightarrow TM$ (the \emph{velocity curve}) and $\frac{%
d^{2}\gamma }{dt^{2}}:[a,b]\rightarrow T^{2}M\subset TTM$ (the \emph{%
acceleration curve}) are the \emph{first order lift} and the \emph{second
order lift} respectively, of the curve $\gamma $. A \emph{second order
Lagrangian} on $M$ is a differentiable map $L^{(2)}:I\!\!R\times
T^{2}M\rightarrow I\!\!R$, where $T^{2}M$ is the second order tangent space
of $M$ (see the Appendix). The \emph{action} of $L^{(2)}$ on $\gamma $ is
given by the formula:%
\begin{equation}
I_{L^{(2)}}(\gamma )=\int_{a}^{b}L^{(2)}\left( t,\gamma (t),\frac{d\gamma }{%
dt}(t),\frac{d^{2}\gamma }{dt^{2}}(t)\right) dt.  \label{action002}
\end{equation}

A second order Lagrangian $L$ is \emph{affine in accelerations} if its
vertical Hessian vanishes; using local coordinates, $%
L(t,x^{i},y^{i},z^{i})=f_{0}(t,x^{i},y^{i})+z^{i}g_{i}(t,x^{i},y^{i})$.
Notice that if $g_{i}=0$, then $L=f_{0}$ is a first order Lagrangian. In
this case $f_{0}$ is obtained projecting $L:T^{2}M\rightarrow I\!\!R$ on $%
f_{0}:TM\rightarrow I\!\!R$, by the natural projection $T^{2}M\rightarrow TM$%
; the degeneration case described in this situation is refined later in the
paper.

It is easy to see that the Lagrangian action $I_{L_{0}}$ of a Lagrangian $%
L_{0}$ is the same as the tangent action $I_{\omega _{0}}$ of the Lagrangian
tangent form $\omega _{0}=L_{0}dt\in \mathcal{X}^{\ast }(I\!\!R\times TM)$.
It worth to remark that $\omega _{0}$ is a closed form only if $%
L_{0}=L_{0}(t)$.

The two actions (\ref{action001}) and (\ref{action002}) are related as
follows.

\begin{pr}
\label{pr P-L}If $\omega \in \mathcal{X}^{\ast }(I\!\!R\times TM)$ is a
tangent form, then there is a second order Lagrangian $L^{(2)}:T^{2}M%
\rightarrow I\!\!R$, affine in accelerations, such that $I_{\omega }=I_{L}$.
\end{pr}

\emph{Proof.} Let $z\in T^{2}M$ and $\gamma :[a,b]\rightarrow M$ be a curve, 
$t^{\prime }\in (a,b)$ and $z=\frac{d^{2}\gamma }{dt^{2}}(t^{\prime })$.
Then we define $L_{\omega }^{(2)}:T^{2}M\rightarrow I\!\!R$,$\ L_{\omega
}^{(2)}(z)=\omega _{\tilde{\gamma}(t)}\left( \frac{d^{2}\gamma }{dt^{2}}%
(t)\right) $. It is easy to see that the actions of $L_{\omega }^{(2)}$ and $%
\omega $ on a curve $\gamma $ have the same form, given by the right side of
the formula (\ref{action001}), thus the conclusion follows. $\Box $

Using coordinates $(x^{i},y^{i},z^{i})$ on $T^{2}M$ (see Appendix), if $%
\omega $ is given by (\ref{eqlocom}), then we have: 
\begin{equation}
L_{\omega }^{(2)}(t,x^{i},y^{i},z^{i})=\omega _{0}(t,x^{i},y^{i})+\omega
_{i}(t,x^{j},y^{j})y^{i}+\bar{\omega}_{i}(t,x^{j},y^{j})z^{i}.
\label{Lagr-ord-2}
\end{equation}

The following result shows that the action of every second order Lagrangian,
affine in accelerations, can be represented as well as an action of a
suitable tangent form.

\begin{pr}
\label{pr L-P}Let $L^{(2)}:I\!\!R\times T^{2}M\rightarrow I\!\!R$ be a
second order Lagrangian affine in accelerations. Then there is a tangent
form $\omega \in \mathcal{X}^{\ast }(I\!\!R\times TM)$ such that $I_{\omega
}=I_{L}$.
\end{pr}

\emph{Proof.} Let us consider a local chart $(U,\varphi )$ on $M$; we define
a locally tangent form $\theta _{U}=\frac{\partial L}{\partial z^{i}}dy^{i}$%
, thus $(\overline{\theta _{U}})_{i}=\frac{\partial L}{\partial z^{i}}$ and $%
(\theta _{U})_{i}=(\theta _{U})_{0}=0$ for this tangent form. Let $%
\{f_{\alpha }\}_{\alpha \in I\!\!N}$ be a partition of unity subordinated to
a locally finite open cover $\{U_{\alpha }\}_{\alpha \in I\!\!N}$ of such
domains of coordinates. Then the tangent form $\theta =\sum\limits_{n\in
I\!\!N}f_{\alpha }\cdot \theta _{U_{\alpha }}$ is a tangent form $\theta \in 
\mathcal{X}^{\ast }(I\!\!R\times TM)$ that has the top component $\bar{\theta%
}_{i}=(\overline{\theta _{U}})_{i}=\frac{\partial L}{\partial z^{i}}$ and $%
\theta =\bar{\theta}_{i}dy^{i}+\theta _{i}dx^{i}$. Since $L^{(2)}$ has the
local expression $L^{(2)}(t,x^{i},y^{i},z^{i})=$ $\bar{\theta}%
_{i}(t,x^{j},y^{j})z^{i}+u(t,x^{j},y^{j})$, one has also $%
L^{(2)}(t,x^{i},y^{i},z^{i})=$ $\bar{\theta}_{i}(t,x^{i},y^{i})z^{i}+\theta
_{i}(t,x^{i},y^{i})y^{i}+(u(t,x^{i},y^{i})-\theta _{i}y^{i})$. Then the
local functions $L_{0}(t,x^{i},y^{i})=u(t,x^{i},y^{i})-\theta _{i}y^{i}$
give a global function $L_{0}:I\!\!R\times TM\rightarrow I\!\!R$ and the
tangent form $\omega =\theta +L_{0}dt$ has the property that $I_{\omega
}=I_{L}$. $\Box $

The actions of tangent forms on curves are related to the well-known actions
of the first and the second order Lagrangians on curves. Let us consider two
points $x$, $y\in M$ and $\gamma _{0}=(x_{0}^{i}(t))$ be a curve joining $x$
and $y$, i.e. $x_{0}^{i}(0)=x$ and $x_{0}^{i}(1)=y$. Let us consider
variations of $\gamma _{0}$, as curves joining $x$ and $y$, having the local
expression $\gamma _{\varepsilon }=(x_{\varepsilon }^{i}(t))$, where $%
x_{\varepsilon }^{i}(t)=x_{0}^{i}(t)+\varepsilon h^{i}(t)$.

In the case of the actions of second order Lagrangians on curves, the
specific variational conditions, impose:

\begin{eqnarray}
h^{i}(a) &=&h^{i}(b)=0,  \label{cond_capete_1} \\
{{{{\frac{dh^{i}}{dt}}}}}(a) &=&{{{{\frac{dh^{i}}{dt}}}}}(b)=0.
\label{cond_capete_2}
\end{eqnarray}

For a second order Lagrangian $L^{(2)}:I\!\!R\times T^{2}M\rightarrow I\!\!R$%
, the extrema curves of the action $I_{L^{(2)}}$ are given by the well-known
Euler-Lagrange equations 
\begin{equation}
\frac{\partial L^{(2)}}{\partial x^{i}}-\frac{d}{dt}\frac{\partial L^{(2)}}{%
\partial y^{i}}+\frac{d^{2}}{dt^{2}}\frac{\partial L^{(2)}}{\partial z^{i}}%
=0.  \label{EL-ord-2}
\end{equation}%
In the particular case of a Lagrangian (\ref{Lagr-ord-2}), the
Euler-Lagrange equations have the form%
\begin{equation}
{{\frac{\partial \omega _{0}}{\partial x^{i}}+\frac{\partial \omega _{j}}{%
\partial x^{i}}\frac{dx_{0}^{j}}{dt}+\frac{\partial \bar{\omega}_{j}}{%
\partial x^{i}}\frac{d^{2}x_{0}^{j}}{dt^{2}}}}-{{{{\frac{d}{dt}(}}\frac{%
\partial \omega _{0}}{\partial y^{i}}}}+{{\frac{\partial \omega _{j}}{%
\partial y^{i}}\frac{dx_{0}^{j}}{dt}}}+\omega _{i}+{{\frac{\partial \bar{%
\omega}_{j}}{\partial y^{i}}\frac{d^{2}x_{0}^{j}}{dt^{2}})}}+{{{{\frac{d^{2}%
}{dt^{2}}}}}}\bar{\omega}_{i}=0.  \label{EL-ord-2+}
\end{equation}

Let us consider $I\!\!R^{2}$ with coordinates $x$ and $y$. The canonical
symplectic form $\alpha =dx\wedge dy$ gives the tangent form $\omega ^{(1)}=%
\dot{x}d\dot{y}-\dot{y}d\dot{x}$ and the second order Lagrangian $L_{0}(t,%
\dot{x},\dot{y},\ddot{x},\ddot{y})=\dot{x}\ddot{y}-\dot{y}\ddot{x}$ on $%
I\!\!R^{2}$; here $(x,y):=(x^{1},x^{2})$; $(\dot{x},\dot{y}):=(y^{1},y^{2})$%
, in the previous notations. This Lagrangian was involved in \cite{LSZ01},
concerning its invariance to the $(2+1)$-Galilean symmetry; the authors
prove in the Appendix that \emph{the general form of a one-particle
Lagrangian which is at most linearly dependent on }$\ddot{x},$\emph{and }$%
\ddot{y}$ \emph{leading to Euler-Lagrange equations of motion which are
covariant with respect to the }$D=2$\emph{\ Galilei group, is given, up to
gauge transformations, by }$L(t,x,y,\dot{x},\dot{y},\ddot{x},\ddot{y})=-k(%
\dot{x}\ddot{y}-\dot{y}\ddot{x})+$ $\frac{m}{2}(\dot{x}^{2}+\dot{y}^{2})$.
This Lagrangian is affine in accelerations, but it can come from two tangent
forms: 
\begin{eqnarray}
\omega _{1} &=&-k(\dot{x}d\dot{y}-\dot{y}d\dot{x})+\frac{m}{2}(\dot{x}dx+%
\dot{y}dy),  \notag \\
\omega _{2} &=&-k(\dot{x}d\dot{y}-\dot{y}d\dot{x})+\frac{m}{2}(\dot{x}^{2}+%
\dot{y}^{2})dt.  \label{om12}
\end{eqnarray}%
In order to put together these two tangent forms, we define below an
equivalence relation, ruled by their action and implicitly by their second
order Lagrangians, affine in accelerations.

\section{Equivalence of tangent forms}

A first order Lagrangian $F:I\!\!R\times TM\rightarrow I\!\!R$ and a tangent
form $\omega \in \mathcal{X}^{\ast }(I\!\!R\times TM)$ give together a
tangent form $\omega ^{\prime }=\omega +dF$. Then

$I_{\omega ^{\prime }}(\gamma )=I_{\omega }(\gamma )+\int_{a}^{b}(\frac{%
\partial F}{\partial t}+\frac{\partial F}{\partial x^{i}}{{\frac{dx^{i}}{dt}}%
}+\frac{\partial F}{\partial y^{i}}{{\frac{dy^{i}}{dt}}})dt=$ $I_{\omega
}(\gamma )+F(x^{i}(b),\frac{dx^{i}}{dt}(b))-$ $F(x^{i}(a),\frac{dx^{i}}{dt}%
(a))$.

According to the variation conditions (\ref{cond_capete_1}) and (\ref%
{cond_capete_2}), it is easy to see that $I_{\omega }$ and $I_{\omega
^{\prime }}$ have the same extrema curves.

Analogous considerations as made in \cite{Leb} for the gauge equivalence of
first order Lagrangians can be transposed for second order Lagrangians (see
for example \cite[Section 4.4]{Mi}). It reads that the second order
Lagrangians $L$ and $L^{\prime }=L+\frac{d}{dt}F$, where $F:I\!\!R\times
TM\rightarrow I\!\!R$, are gauge equivalent, i.e. they have the same extrema
curves. Here $\frac{d}{dt}F$ stands for $L_{dF}$, the second order
Lagrangian associated with the tangent form $dF$. The analogous gauge form
for actions of the corresponding tangent forms, reads that the tangent forms 
$\omega $ and $\omega ^{\prime }=\omega +dF$ have the same extrema curves.

We notice that the Lagrangians given by \cite[formula (11)]{CMR02} or \cite[
formula (34)]{CMR} are gauge equivalent, but they are studied without using
this fact.

Let us consider the submodule $\mathcal{G}\subset \mathcal{X}^{\ast
}(I\!\!R\times TM)$ generated (as a sheaf) by the local differential forms $%
\{\delta x^{i}=dx^{i}-y^{i}dt\}_{i=\overline{1,m}}$. A differential form $%
\eta \in \mathcal{G}$ iff it has the local expression $\eta
=a_{i}(t,x^{j},y^{j})\delta x^{i}$. It is easy to see that any form $\eta
\in \mathcal{G}$ vanishes along the (second order) lift of a curve on $M$.
Thus for any tangent form $\omega \in \mathcal{X}^{\ast }(I\!\!R\times TM)$,
the tangent forms $\omega $ and $\omega ^{\prime }=\omega +\eta $ have the
same extrema curves (see \cite{Ma03} for other implications concerning the
module $\mathcal{G}$).

We say that:

\begin{description}
\item two tangent forms $\omega ,\omega ^{\prime }\in \mathcal{X}^{\ast
}(I\!\!R\times M)$ are \emph{equivalent} if there is an $F\in \mathcal{X}%
^{\ast }(I\!\!R\times M)$ such that $\omega ^{\prime }-\omega -dF\in $ $%
\mathcal{G}$;

\item two second order Lagrangians $L^{\prime }$ and $L$ are \emph{gauge
equivalent} if there is an $F\in \mathcal{X}^{\ast }(I\!\!R\times TM)$ such
that $L^{\prime }-L=\frac{d}{dt}F$.
\end{description}

It is easy to see that two equivalent tangent forms have the same extrema
curves. Analogously, two second order Lagrangians $L^{\prime }$ and $L$ that
are gauge equivalents have the same extrema curves.

\begin{pr}
\label{preqg}Two tangent forms $\omega ^{\prime }$ and $\omega $ are
equivalent iff the corresponding second order Lagrangians $L_{\omega
^{\prime }}$ and $L_{\omega }$ are gauge equivalent.
\end{pr}

\emph{Proof.} Let $\omega ^{\prime }=\bar{\omega}_{i}^{\prime }dy^{i}+\omega
_{i}^{\prime }dx^{i}+\omega _{0}^{\prime }$ and $\omega =\bar{\omega}%
_{i}dy^{i}+\omega _{i}dx^{i}+\omega _{0}$ be equivalent. Thus there is $F\in 
\mathcal{F}(I\!\!R\times M)$ such that $\omega ^{\prime }-\omega -dF=\eta
_{i}(dx^{i}-y^{i}dt)$. It is easy to see that $L_{\omega ^{\prime
}}-L_{\omega }=\frac{d}{dt}F$, thus $L_{\omega ^{\prime }}$ and $L_{\omega }$
are gauge equivalent. Conversely, let us suppose that $L_{\omega ^{\prime }}$
and $L_{\omega }$ are gauge equivalent, thus $L_{\omega ^{\prime
}}-L_{\omega }=\frac{d}{dt}F$. Then $\bar{\omega}_{i}^{\prime }=\bar{\omega}%
_{i}+\frac{\partial F}{\partial y^{i}}$ and $\omega _{i}^{\prime
}y^{i}+\omega _{0}^{\prime }=(\frac{\partial F}{\partial y^{i}}+\omega
_{i})y^{i}+\frac{\partial F}{\partial t}+\omega _{0}$. It follows that $%
\omega ^{\prime }-\omega -dF=(\omega _{i}^{\prime }-\omega _{i}-\frac{%
\partial F}{\partial x^{i}})(dx^{i}-y^{i}dt)\in \mathcal{G}$, thus $\omega
^{\prime }$ and $\omega $ are equivalent. $\Box $

It follows that the property of the above Proposition can be used as a
definition of equivalent tangent forms.

\begin{cor}
\label{corech}If two tangent forms correspond to the same second order
Lagrangian affine in accelerations, then they are equivalent.
\end{cor}

The Poincar\'{e}-Cartan form $\theta _{L}=Ldt+\frac{\partial L}{\partial
y^{i}}\delta x^{i}$ of a first order Lagrangian $L$ is obviously equivalent
to the canonical Lagrangian form $Ldt$ and both correspond to the same
Lagrangian $L$, seen of second order by $T^{2}M\rightarrow TM\rightarrow
I\!\!R$.

There are two possibilities to associate a tangent form to a pointed
Lagrangian $L(t,x^{i},y^{i})=y^{i}\nu _{i}$: $\omega _{1}=\nu _{i}dx^{i}$
and $\omega _{2}=Ldt$ respectively. The first is pure and the second is a
Lagrangian one, but they have the same action on curves, that given by the
action of the same Lagrangian. It is easy to see that $d\omega _{1}-d\omega
_{2}=0$ iff $\nu ^{i}=0$; thus $\omega _{1}$ and $\omega _{2}$ are not
differential equivalent (i.e. $\omega _{1}-\omega _{2}=dF$) for $L\neq 0$.
Thus there are tangent forms that are not differential equivalent (i.e.
their difference is not a exact differential), but equivalent.

Every tangent form $\omega $ of the form (\ref{eqlocom}) is locally
equivalent to the local tangent form $\omega ^{\prime }=(\omega
_{0}+y^{i}\omega _{i})dt+\bar{\omega}_{i}dy^{i}$, since $\omega -\omega
^{\prime }=\omega _{i}\delta x^{i}$. But, in general $\omega ^{\prime }$ is
not a global tangent form.

The tangent forms $\omega _{1}=-k(\dot{x}d\dot{y}-\dot{y}d\dot{x})+$ $\frac{m%
}{2}(\dot{x}dx+\dot{y}dy)$ and $\omega _{2}=-k(\dot{x}d\dot{y}-\dot{y}d\dot{x%
})+$ $\frac{m}{2}(\dot{x}^{2}+\dot{y}^{2})$ considered previously are
equivalent, since $\omega _{1}-\omega _{2}=\frac{m}{2}(\dot{x}(dx-\dot{x}dt)+%
\dot{y}(dy-\dot{y}dt))$. Let us consider below two other situations.

1) Considering the canonical symplectic form in $I\!\!R^{2}$ given by $%
(\varepsilon _{ij})=\left( 
\begin{array}{cc}
0 & 1 \\ 
-1 & 0%
\end{array}%
\right) $, we obtain the pure tangent form $\omega _{1}=-k\varepsilon
_{ij}y^{i}dy^{j}$, or $\omega _{1}=-k\varepsilon _{ij}\dot{x}^{i}d\dot{x}%
^{j} $ where $k$ is a non-null constant, that corresponds to the second
order Lagrangian $L_{0}(x^{i},y^{i},z^{i})=-k\varepsilon _{ij}y^{i}z^{j}$,
or $L_{0}(x^{i},\dot{x}^{i},\ddot{x}^{j})=-k\varepsilon _{ij}\dot{x}^{i}%
\ddot{x}^{j}$.

The Lagrangian $L(x^{i},y^{i},z^{i})=\frac{my^{i}y^{j}\delta _{ij}}{2}%
-k\varepsilon _{ij}y^{i}z^{j}$, or $L(x^{i},\dot{x}^{i},\ddot{x}^{j})=\frac{m%
\dot{x}^{i}\dot{x}^{j}\delta _{ij}}{2}-k\varepsilon _{ij}\dot{x}^{i}\ddot{x}%
^{j}$ was considered in \cite{LSZ01, LSZ02, Ac}.

In order to obtain a tangent form we have two possibilities: $\omega =$ $%
my^{i}\delta _{ij}dx^{j}-k\varepsilon _{ij}y^{i}dy^{j}$ and $\omega ^{\prime
}=$ $\frac{my^{i}y^{j}\delta _{ij}}{2}dt-k\varepsilon _{ij}y^{i}dy^{j}$; the
first is pure and the second is a mixed one.

2) The Lagrangian $L(x^{i},y^{i},z^{i})=-m\left\Vert y^{(1)}\right\Vert +%
\frac{\varepsilon _{ij}y^{i}z^{j}}{\left\Vert y^{(1)}\right\Vert ^{3}}$,
where $\left\Vert y^{(1)}\right\Vert =\sqrt{\frac{y^{i}y^{j}\delta _{ij}}{2}}
$ or $L(x^{i},\dot{x}^{i},\ddot{x}^{j})=-m\left\Vert \dot{x}\right\Vert +%
\frac{\varepsilon _{ij}\dot{x}^{i}\ddot{x}^{j}}{\left\Vert \dot{x}%
\right\Vert ^{3}}$, where $\left\Vert \dot{x}\right\Vert =\sqrt{\frac{m\dot{x%
}^{i}\dot{x}^{j}\delta _{ij}}{2}}$ was considered in \cite{Ma02}. The two
tangent forms, one pure and one mixed, can also be considered: $\omega =-%
\frac{my^{i}\delta _{ij}}{\left\Vert y^{(1)}\right\Vert }dx^{j}+\frac{%
\varepsilon _{ij}y^{i}}{\left\Vert y^{(1)}\right\Vert ^{3}}dy^{j}$, and $%
\omega ^{\prime }=-m\left\Vert y^{(1)}\right\Vert dt+\frac{\varepsilon
_{ij}y^{i}}{\left\Vert y^{(1)}\right\Vert ^{3}}dy^{j}$.

Unlike the first example, in the second example the tangent form $\omega
^{\prime }$ is not differentiable in the points where $(y^{i}=0)$.

\subsection{Controlled and higher order tangent forms}

We consider below controlled tangent form and, in particular, higher order
tangent forms. The controlled top derivative, defined also in this
subsection, is used in the next subsection in an accurate study of the
Euler-Lagrange equation of a tangent form.

Let $\pi _{E}:E\rightarrow M$ be a fiber manifold (i.e. a surjective
submersion) that factorize as a composition $\pi _{E}=\pi _{TM}\circ \pi
_{E}^{\prime }$ a fibered manifolds $\pi _{E}^{\prime }:E\rightarrow TM$ and
the canonical projection $\pi _{TM}:TM\rightarrow M$.

A \emph{controlled tangent form} on $E$ is a fibered map $\omega
:I\!\!R\times E\rightarrow T^{\ast }TM$ over the base $TM$. A \emph{%
controlled top tangent form} on $E$ is a bundle map $\bar{\omega}%
:I\!\!R\times E\rightarrow \pi _{TM}^{\ast }T^{\ast }M\cong V^{\ast }TM$
over the base $TM$.

Obviously a controlled tangent form $\omega $ as above gives rise to a top
tangent form $\bar{\omega}=J^{\ast }\circ \omega $.

Let us consider some local coordinates, adapted to the fibered structures: $%
(x^{i})$ on $M$, $(x^{i},y^{j})$ on $TM$ and $(x^{i},y^{j},u^{\bar{\alpha}%
})=(x^{i},u^{\alpha })$ on $E$. A controlled tangent form $\omega $ has the
expression $(t,x^{i},u^{\alpha })\rightarrow (x^{i},y^{i},\omega
_{i}(t,x^{i},u^{\alpha }),\bar{\omega}_{i}(t,x^{i},u^{\alpha }))$; its
expression on fibers is $\omega =\omega _{i}(t,x^{i},u^{\alpha })dx^{i}+\bar{%
\omega}_{i}(t,x^{i},u^{\alpha })dy^{i}$. Considering the (local)
differential operator $\frac{d}{dt}=\frac{\partial }{\partial t}+y^{i}\frac{%
\partial }{\partial x^{i}}+v^{\alpha }\frac{\partial }{\partial u^{\alpha }}$
on the sheaf of local real functions on $TE$ having the same domain of
definition, we say that 
\begin{equation}
\mathcal{E}_{\omega }^{\prime }(t,x^{i},y^{i},u^{\alpha },v^{\alpha
})=(\omega _{i}-\frac{d}{dt}\bar{\omega}_{i})dx^{i}  \label{eqlagrder}
\end{equation}%
is the \emph{Lagrange controlled top derivative} of $\omega $.

\begin{pr}
\label{prdefconder}The Lagrange controlled top derivative of $\omega $ is a
global fibered map $\mathcal{E}_{\omega }^{\prime }:I\!\!R\times
TE\rightarrow T^{\ast }M$, over $M$.
\end{pr}

\emph{Proof.} Considering the local expressions $\mathcal{E}_{\omega
}^{\prime }(t,x^{i},y^{i},u^{\alpha },v^{\alpha })=[\omega _{i}-(\frac{%
\partial \bar{\omega}_{i}}{\partial x^{j}}y^{j}+\frac{\partial \bar{\omega}%
_{i}}{\partial u^{\alpha }}v^{\alpha }+\frac{\partial \bar{\omega}_{i}}{%
\partial t})]dx^{i}$, we have to prove that the definition does not depend
on coordinates. If $\{x^{i^{\prime }},y^{i^{\prime }},u^{\alpha ^{\prime
}},v^{\alpha ^{\prime }}\}$ is another set of coordinates, on the
intersection domain we have the rules $\bar{\omega}_{i}=\frac{\partial
x^{i^{\prime }}}{\partial x^{i}}\bar{\omega}_{i^{\prime }}$ and $\omega _{i}=%
\frac{\partial y^{i^{\prime }}}{\partial x^{i}}\bar{\omega}_{i^{\prime }}+%
\frac{\partial x^{i^{\prime }}}{\partial x^{i}}\omega _{i^{\prime }}$
respectively. Analogously, a controlled top tangent form $\bar{\omega}$ has
the expression $\bar{\omega}=\bar{\omega}_{i}dx^{i}$ and $\bar{\omega}_{i}=%
\frac{\partial x^{i^{\prime }}}{\partial x^{i}}\bar{\omega}_{i^{\prime }}$.

We have $\omega _{i}-(\frac{\partial \bar{\omega}_{i}}{\partial x^{j}}y^{j}+%
\frac{\partial \bar{\omega}_{i}}{\partial u^{\alpha }}v^{\alpha }+\frac{%
\partial \bar{\omega}_{i}}{\partial t})=\frac{\partial y^{i^{\prime }}}{%
\partial x^{i}}\bar{\omega}_{i^{\prime }}+\frac{\partial x^{i^{\prime }}}{%
\partial x^{i}}\omega _{i^{\prime }}-$

$y^{j}\frac{\partial ^{2}x^{i^{\prime }}}{\partial x^{j}\partial x^{i}}\bar{%
\omega}_{i^{\prime }}-y^{j}\frac{\partial x^{i^{\prime }}}{\partial x^{i}}(%
\frac{\partial u^{\alpha ^{\prime }}}{\partial x^{j}}\frac{\partial \bar{%
\omega}_{i^{\prime }}}{\partial u^{\alpha ^{\prime }}}+\frac{\partial
x^{j^{\prime }}}{\partial x^{j}}\frac{\partial \bar{\omega}_{i^{\prime }}}{%
\partial x^{j^{\prime }}})-\frac{\partial x^{i^{\prime }}}{\partial x^{i}}%
\frac{\partial \bar{\omega}_{i^{\prime }}}{\partial u^{\alpha ^{\prime }}}%
\frac{\partial u^{\alpha ^{\prime }}}{\partial u^{\alpha }}v^{\alpha }-\frac{%
\partial x^{i^{\prime }}}{\partial x^{i}}\frac{\partial \bar{\omega}%
_{i^{\prime }}}{\partial t}=$

$\frac{\partial x^{i^{\prime }}}{\partial x^{i}}(\omega _{i^{\prime }}-\frac{%
\partial \bar{\omega}_{i^{\prime }}}{\partial x^{j^{\prime }}}y^{j^{\prime
}}-\frac{\partial \bar{\omega}_{i^{\prime }}}{\partial u^{\alpha ^{\prime }}}%
v^{\alpha ^{\prime }}-\frac{\partial \bar{\omega}_{i^{\prime }}}{\partial t}%
) $, thus the conclusion follows. $\Box $

It is easy to see that in the case $E=TM$, we obtain that a controlled (top)
tangent form is just a (top) tangent form. The Lagrange controlled
derivative of a tangent form $\omega :\mathcal{E}_{\omega }^{\prime
}:I\!\!R\times TTM\rightarrow T^{\ast }TM$ restricts to the Lagrange
derivative of $\omega $: $\mathcal{E}_{\omega }^{\prime }:I\!\!R\times
T^{2}M\rightarrow T^{\ast }TM$ given above by formula (\ref{eqlagrder}); it
follows using the local expression of the inclusion $T^{2}M\subset TTM$ (see
the Appendix).

We define:

\begin{description}
\item a $k$\emph{-order tangent form} as a controlled tangent form $\omega
:I\!\!R\times T^{k}M\rightarrow T^{\ast }TM$ and

\item a $k$\emph{-order top tangent form as }a bundle map $\bar{\omega}%
:I\!\!R\times T^{k}M\rightarrow \pi _{TM}^{\ast }T^{\ast }M$.
\end{description}

As in the general case, a $k$-order tangent form $\omega $ gives a $k$-order
top tangent form $\bar{\omega}=J^{\ast }\circ \omega $.

Let us define now the \emph{Lagrange top derivative} $\mathcal{E}_{\omega
}^{(k+1)}$ of a $k$-order tangent form $\omega :I\!\!R\times
T^{k}M\rightarrow T^{\ast }TM$. The Lagrange controlled top derivative of $%
\omega $ is $\mathcal{E}_{\omega }^{(k+1)}:I\!\!R\times TT^{k}M\rightarrow
\pi _{TM}^{\ast }T^{\ast }M$ restricts to the Lagrange derivative of $\omega 
$: $\mathcal{E}_{\omega }^{(k+1)}:I\!\!R\times T^{k+1}M\rightarrow \pi
_{TM}^{\ast }T^{\ast }M$, according to the inclusion $T^{k+1}M\subset
TT^{k}M $. Using local coordinates, the inclusion has the form $%
(x^{i},y^{i},\ldots ,w^{i},\tilde{w}^{i})\rightarrow $ $(x^{i},y^{i},\ldots
,w^{i},y^{i},\ldots , $ $w^{i},$ $\tilde{w}^{i})$, $\mathcal{E}_{\omega
}^{(k+1)}:I\!\!R\times \pi _{TM}^{\ast }TT^{k}M=TT^{k}M\rightarrow \pi
_{TM}^{\ast }T^{\ast }M$ has the local expression

\begin{center}
$\mathcal{E}_{\omega }^{\prime }(t,x^{i},y^{i},\ldots
,w^{i},X^{i},Y^{i},\ldots ,W^{i})=[\omega _{i}-(\frac{\partial \bar{\omega}%
_{i}}{\partial t}+\frac{\partial \bar{\omega}_{i}}{\partial x^{j}}X^{j}+%
\frac{\partial \bar{\omega}_{i}}{\partial y^{j}}Y^{j}+\cdots +\frac{\partial 
\bar{\omega}_{i}}{\partial w^{j}}W^{j})]dx^{i}$
\end{center}

\noindent and the restriction to $I\!\!R\times \pi _{TM}^{\ast }T^{k+1}M$ is 
\begin{eqnarray*}
\mathcal{E}_{\omega }^{(k+1)} &:&I\!\!R\times T^{k+1}M\rightarrow T^{\ast }M,%
\mathcal{E}_{\omega }^{(k+1)}(t,x^{i},y^{i},\ldots ,z^{i},w^{i},\tilde{w}%
^{i})= \\
&&[\omega _{i}-(\frac{\partial \bar{\omega}_{i}}{\partial t}+\frac{\partial 
\bar{\omega}_{i}}{\partial x^{j}}y^{j}+\cdots +\frac{\partial \bar{\omega}%
_{i}}{\partial z^{i}}w^{i}+\frac{\partial \bar{\omega}_{i}}{\partial w^{i}}%
\tilde{w}^{i})]dx^{i},
\end{eqnarray*}%
or 
\begin{equation}
\mathcal{E}_{\omega }^{(k+1)}=[\omega _{i}-\frac{d}{dt}\bar{\omega}%
_{i}]dx^{i},  \label{eqdefek}
\end{equation}%
where $\frac{d}{dt}$ is the local operator given by $\frac{d}{dt}=\frac{%
\partial }{\partial t}+y^{j}\frac{\partial }{\partial x^{j}}+\cdots +w^{i}%
\frac{\partial }{\partial z^{i}}+\tilde{w}^{i}\frac{\partial }{\partial w^{i}%
}$.

The first order Lagrange top derivative of a tangent form $\omega =\omega
_{i}dx^{i}+\bar{\omega}_{i}dy^{i}$ is the second order top tangent form $%
\mathcal{E}_{\omega }^{(2)}=\mathcal{E}_{\omega }^{\prime }:I\!\!R\times \pi
_{TM}^{\ast }T^{2}M\rightarrow \pi _{TM}^{\ast }T^{\ast }M$ given using (\ref%
{eqdefek}).

In the case $k=3$, the local operator $\frac{d}{dt}$ is given by $\frac{d}{dt%
}=\frac{\partial }{\partial t}+y^{j}\frac{\partial }{\partial x^{j}}+z^{i}%
\frac{\partial }{\partial y^{i}}+w^{i}\frac{\partial }{\partial z^{i}}+%
\tilde{w}^{i}\frac{\partial }{\partial w^{i}}$. In the case when $\bar{\omega%
}=\bar{\omega}_{i}dx^{i}$ has the order $2$, then $\frac{\partial \bar{\omega%
}_{i}}{\partial w^{i}}=0$ and the third order Lagrange top derivative $%
\mathcal{E}_{\omega }^{(3)}$ has the order at most $3$, as $\omega $. This
is the case below when the Euler-Lagrange top form of a tangent form has the
third order.

\section{The Euler-Lagrange equation as a top tangent form}

Any second order Lagrangian $L:I\!\!R\times T^{2}M\rightarrow I\!\!R$ gives
rise to an at most forth order top tangent form $\mathcal{E}_{i}dx^{i}$,
that we call the \emph{Euler-Lagrange top tangent form} of $L$, where

\begin{equation}
\mathcal{E}_{i}=\frac{\partial L}{\partial x^{i}}-\frac{d}{dt}\frac{\partial
L}{\partial y^{i}}+\frac{d^{2}}{dt^{2}}\frac{\partial L}{\partial z^{i}},
\label{eqdefELgen}
\end{equation}%
$\frac{d}{dt}=\frac{\partial }{\partial t}+y^{j}\frac{\partial }{\partial
x^{j}}+z^{j}\frac{\partial }{\partial y^{j}}+w^{j}\frac{\partial }{\partial
z^{j}}+\tilde{w}^{j}\frac{\partial }{\partial w^{j}}$ and $%
(x^{i},y^{i},z^{i},w^{i},\tilde{w}^{i})$ are the canonical local coordinates
on $T^{4}M$ induced by the local coordinates $(x^{i})$ on $M$. In the case
when a second order Lagrangian $L_{\omega }$ is affine in accelerations and
it is associated with a tangent form $\omega $, its local formula is given
as in formula (\ref{Lagr-ord-2}) with $L^{(2)}=L_{\omega }$. We say that the
Euler-Lagrange top tangent form $\mathcal{E}=\mathcal{E}_{\omega }$ of $%
L_{\omega }$ is the \emph{Euler-Lagrange top tangent form} of $\omega $.
Specifically, if $\omega $ is a tangent form given by formula (\ref{eqlocom}%
), then $L_{\omega }(t,x^{i},y^{i},z^{i})=\omega _{0}+y^{i}\omega _{i}+z^{i}%
\bar{\omega}_{i}$, thus 
\begin{equation}
\mathcal{E}_{i}={{\frac{\partial \omega _{0}}{\partial x^{i}}+\frac{\partial
\omega _{j}}{\partial x^{i}}y^{j}+\frac{\partial \bar{\omega}_{j}}{\partial
x^{i}}z^{j}}}-{{{{\frac{d}{dt}(}}\frac{\partial \omega _{0}}{\partial y^{i}}}%
}+{{\frac{\partial \omega _{j}}{\partial y^{i}}y^{j}}}+\omega _{i}+{{\frac{%
\partial \bar{\omega}_{j}}{\partial y^{i}}z^{j})}}+{{{{\frac{d^{2}}{dt^{2}}}}%
}}\bar{\omega}_{i},  \label{eqdefEL}
\end{equation}%
where $\frac{d}{dt}=\frac{\partial }{\partial t}+y^{j}\frac{\partial }{%
\partial x^{j}}+z^{j}\frac{\partial }{\partial y^{j}}+w^{j}\frac{\partial }{%
\partial z^{j}}$, since $\frac{\partial L}{\partial z^{i}}=\bar{\omega}%
_{i}(t,x^{i},y^{i},z^{i})$ and the forth order coordinates $\left( \tilde{w}%
^{i}\right) $ are not involved. Thus the top tangent form $\mathcal{E}$ is
at most third order in this case.

We prove below that the Euler-Lagrange top tangent form can be obtained
using two second order tangent forms.

\begin{pr}
\label{pr-ostr}Let $\omega $ be a (first order) tangent form such that the
Euler-Lagrange top tangent form $\mathcal{E}_{\omega }$ is of third order.
Then the following assertions holds true.

\begin{enumerate}
\item If $\Omega $ is a first or a second order tangent form such that such
that $\bar{\Omega}=\bar{\omega}$, then there is a second or a third order
tangent form $\Phi $, uniquely determined by the conditions that the
Lagrange top derivative of $\Omega $ is $\bar{\Phi}$ and the Lagrange top
derivative of $\Phi $ is the Euler-Lagrange top tangent form $\mathcal{E}%
_{\omega }$.

\item There are two second order tangent forms $\Omega $ and $\Phi $ such
that $\bar{\Omega}=\bar{\omega}$, the Lagrange top derivative of $\Omega $
is $\bar{\Phi}$ and the Lagrange top derivative of $\Phi $ is $\mathcal{E}%
_{\omega }$.
\end{enumerate}
\end{pr}

\emph{Proof.} 1. The conditions on $\Phi $ read $\bar{\Phi}_{i}=\Omega _{i}-%
\frac{d}{dt}\bar{\Omega}_{i}=$ $\Omega _{i}-\frac{d}{dt}\bar{\omega}_{i}$
and $\Phi _{i}=\frac{d}{dt}\bar{\Phi}_{i}+\mathcal{E}_{\omega }$, thus $%
\Omega $ is uniquely given by these conditions. If $\Omega $ is of first
order then $\bar{\Phi}$ is of second order, thus $\Phi $ is of second or of
third order. If $\Omega $ is of second order, then $\bar{\Phi}$ is of second
or third order, thus $\Phi $ is of second or third order.

2. Let us denote $\omega =\omega _{0}dt+\omega _{i}dx^{i}+\bar{\omega}%
_{i}dy^{i}$ and let $L=$ $\omega _{0}+\omega _{i}y^{i}+\bar{\omega}_{i}z^{i}$
be the associated two order Lagrangian affine in accelerations. We consider $%
\Omega =\Omega _{i}dx^{i}+\bar{\Omega}_{i}dy^{i}=\left( \frac{\partial L}{%
\partial y^{i}}-\omega _{i}\right) dx^{i}+\frac{\partial L}{\partial z^{i}}%
dy^{i}=$ ${{{(}\frac{\partial \omega _{0}}{\partial y^{i}}}}+{{\frac{%
\partial \omega _{j}}{\partial y^{i}}y^{j}}}+{{\frac{\partial \bar{\omega}%
_{j}}{\partial y^{i}}z^{j})dx}}^{i}+\bar{\omega}_{i}dy^{i}$. We have $\frac{%
\partial }{\partial y^{i}}=\frac{\partial y^{i^{\prime }}}{\partial z^{i}}%
\frac{\partial }{\partial z^{i}}+2\frac{\partial x^{i^{\prime }}}{\partial
x^{i}}\frac{\partial }{\partial y^{i}}$ and $\frac{\partial }{\partial z^{i}}%
=\frac{\partial x^{i^{\prime }}}{\partial z^{i}}\frac{\partial }{\partial
z^{i}}$. It follows that 
\begin{eqnarray}
\bar{\Omega}_{i} &=&\frac{\partial x^{i^{\prime }}}{\partial x^{i}}\bar{%
\Omega}_{i^{\prime }},  \label{eqchangeoom} \\
\Omega _{i} &=&\frac{\partial y^{i^{\prime }}}{\partial x^{i}}\bar{\Omega}%
_{i^{\prime }}+\frac{\partial x^{i^{\prime }}}{\partial x^{i}}\Omega
_{i^{\prime }},  \notag
\end{eqnarray}%
thus the local $1$--forms $\Omega =\Omega _{i}dx^{i}+\bar{\Omega}_{i}dy^{i}$
are the restrictions of a (global) second order tangent form $\Omega
:I\!\!R\times T^{2}M\rightarrow T^{\ast }TM$.

According to 1., $\bar{\Phi}_{i}=\Omega _{i}-\frac{d}{dt}\bar{\omega}_{i}=$ $%
\frac{\partial L}{\partial y^{i}}-\omega _{i}-\frac{d}{dt}\frac{\partial L}{%
\partial z^{i}}$ and $\Phi _{i}=\frac{d}{dt}\bar{\Phi}_{i}+\mathcal{E}%
_{\omega }=$ $\frac{\partial L}{\partial x^{i}}-\frac{d}{dt}\omega _{i}$,
thus $\Phi $ is of second order. $\Box $

We say that the second order tangent form $\Omega $ constructed in 2. of
Proposition \ref{pr-ostr} is an \emph{Ostrogradski tangent form} of $\omega $%
.

The above construction is related to a general approach, related to the
classical Ostrogradski theory.

Let $L:I\!\!R\times T^{2}M\rightarrow I\!\!R$ be a second order Lagrangian.
There is a top tangent form $\bar{\omega}=\frac{\partial L}{\partial z^{i}}%
dx^{i}$ associated with this Lagrangian and having an order at most $2$. Let
us suppose that $\omega $ is a first or second order tangent form $\omega $
such that its top tangent form is $\bar{\omega}$, i.e. $\omega =\omega
_{i}dx^{i}+\frac{\partial L}{\partial z^{i}}dy^{i}$. One can consider for
example $\omega =\frac{1}{2}\frac{\partial L}{\partial y^{i}}dx^{i}+\frac{%
\partial L}{\partial z^{i}}dy^{i}$. Then the formula $\Omega =\Omega
_{i}dx^{i}+\bar{\Omega}_{i}dx^{i}=\left( \frac{\partial L}{\partial y^{i}}%
-\omega _{i}\right) dx^{i}+\frac{\partial L}{\partial z^{i}}dx^{i}$ defines
a tangent form of order at most $2$.

Then $\eta :$ $I\!\!R\times T^{2}M\rightarrow T^{\ast }TM$, $\eta
(t,x^{i},y^{i},z^{i})=\left( \frac{\partial L}{\partial y^{i}}-\omega
_{i}\right) dx^{i}+\frac{\partial L}{\partial z^{i}}dy^{i}$ is a second
order tangent form. The Lagrange top derivative of $\eta $ is $\mathcal{E}%
_{\eta }^{(3)}:I\!\!R\times T^{2}M\rightarrow T^{\ast }M$, $\mathcal{E}%
_{\eta }^{\prime \prime }=\left( \frac{\partial L}{\partial y^{i}}-\omega
_{i}-\frac{d}{dt}\frac{\partial L}{\partial z^{i}}\right) dx^{i}$. Usually $%
\left( \frac{\partial L}{\partial y^{i}}-\omega _{i}-\frac{d}{dt}\frac{%
\partial L}{\partial z^{i}}\right) $ is denoted by $p_{i}$.

Since $\mathcal{E=}\frac{\partial L}{\partial x^{i}}-\frac{d}{dt}\frac{%
\partial L}{\partial y^{i}}+\frac{d^{2}}{dt^{2}}\frac{\partial L}{\partial
z^{i}}=$ $\frac{\partial L}{\partial x^{i}}-\frac{d}{dt}\omega _{i}-\frac{d}{%
dt}\left( \frac{\partial L}{\partial y^{i}}-\omega _{i}-\frac{d}{dt}\frac{%
\partial L}{\partial z^{i}}\right) $, it follows that $\mu =\left( \frac{%
\partial L}{\partial x^{i}}-\frac{d}{dt}\omega _{i}\right) dx^{i}+\left( 
\frac{\partial L}{\partial y^{i}}-\omega _{i}-\frac{d}{dt}\frac{\partial L}{%
\partial z^{i}}\right) dy^{i}$ is at most third order tangent form and its
Lagrange top derivative is $\mathcal{E}_{\mu }^{(4)}=\mathcal{E}_{\omega }$,
the Euler-Lagrange top tangent form of $\omega $. This algorithm can produce
tangent forms for a second order Lagrangian, taking a suitable tangent form $%
\omega $.

A $k$-order semi-spray $S:I\!\!R\times T^{k}M\rightarrow T^{k+1}M$ is a
section $(t,x^{(k)})\overset{\bar{S}}{\rightarrow }{}(t,S(t,x^{(k)}))$ of
the affine bundle $I\!\!R\times T^{k+1}M\rightarrow I\!\!R\times T^{k}M$,
obtained as a product of the affine bundle $T^{k+1}M\rightarrow T^{k}M$ and
the identity $I\!\!R\rightarrow I\!\!R$.

Let $\mathcal{E}_{\omega }:I\!\!R\times T^{3}M\rightarrow T^{\ast }M$ be the
Euler-Lagrange top tangent form of a tangent form $\omega $. We say that a
second order semi-spray $\bar{S}:I\!\!R\times T^{2}M\rightarrow I\!\!R\times
T^{3}M$ is adapted to the tangent form $\omega $ if $\mathcal{E}_{\omega
}\circ \bar{S}=0$. The local expressions of $\bar{S}$ and $\mathcal{E}%
_{\omega }$ are $(t,x^{i},y^{i},z^{i})\overset{\bar{S}}{\rightarrow }%
{}(t,x^{i},y^{i},z^{i},S^{i}(t,x^{i},y^{i},z^{i}))$ and $\mathcal{E}_{\omega
}=\mathcal{E}_{i}dx^{i}$ respectively, with $\mathcal{E}_{i}$ given by (\ref%
{eqdefEL}). Denoting $h_{ij}=\frac{\partial \bar{\omega}_{i}}{\partial y^{j}}%
-\frac{\partial \bar{\omega}_{j}}{\partial y^{i}}$, then there are local
functions $f_{i}(t,x^{i},y^{i},z^{i})$ such that 
\begin{equation*}
\mathcal{E}%
_{i}(t,x^{i},y^{i},z^{i},w^{i})=h_{ij}w^{j}+f_{i}(t,x^{i},y^{i},z^{i}).
\end{equation*}

More precisely:

$\mathcal{E}_{i}={{\frac{\partial \omega _{0}}{\partial x^{i}}}}-{{\frac{%
\partial ^{2}\omega _{0}}{\partial t\partial y^{i}}-{\frac{\partial
^{2}\omega _{0}}{\partial x^{j}\partial y^{i}}}y^{j}}}+({{\frac{\partial
\omega _{j}}{\partial x^{i}}}}-{{\frac{\partial ^{2}\omega _{j}}{\partial
t\partial y^{i}}-\frac{\partial \omega _{j}}{\partial x^{j}\partial y^{i}}-}}
$

${{\frac{\partial \omega _{j}}{\partial y^{j}\partial y^{i}})y^{j}}}-$ ${{{{%
\frac{\partial \omega _{i}}{\partial t}-\frac{\partial \omega _{i}}{\partial
x^{j}}y^{j}}}}}+$ $\frac{\partial ^{2}\bar{\omega}_{i}}{\partial t^{2}}+$ $%
\frac{\partial ^{2}\bar{\omega}_{i}}{\partial x^{j}\partial t}y^{j}{+}$ ${(}%
\frac{\partial ^{2}\bar{\omega}_{i}}{\partial t\partial x^{j}}+$ $\frac{%
\partial ^{2}\bar{\omega}_{i}}{\partial x^{k}\partial x^{j}}y^{k})y^{j}{-}$

${{-{\frac{\partial ^{2}\omega _{0}}{\partial y^{j}\partial y^{i}}}}}z^{j}{-}%
\frac{\partial \omega _{j}}{\partial y^{i}}z^{j}$ ${{{{{-}\frac{\partial
\omega _{i}}{\partial y^{j}}}}}}z^{j}+{{\frac{\partial \bar{\omega}_{j}}{%
\partial x^{i}}}}z^{j}+(2\frac{\partial ^{2}\bar{\omega}_{i}}{\partial
x^{k}\partial y^{j}}-$ ${{\frac{\partial \bar{\omega}_{j}}{\partial
x^{k}\partial y^{i}})y^{k}}}z^{j}{+}$

$+(2\frac{\partial ^{2}\bar{\omega}_{i}}{\partial y^{j}\partial t}{{-\frac{%
\partial ^{2}\bar{\omega}_{j}}{\partial t\partial y^{i}}}})z^{j}{+{\frac{%
\partial \bar{\omega}_{i}}{\partial x^{j}}z^{j}+}}$ $(\frac{\partial ^{2}%
\bar{\omega}_{i}}{\partial y^{k}\partial y^{j}}-{{\frac{\partial \bar{\omega}%
_{j}}{\partial y^{k}\partial y^{i}}}})z^{k}z^{j}{+}$ $\left( {{\frac{%
\partial \bar{\omega}_{i}}{\partial y^{j}}-\frac{\partial \bar{\omega}_{j}}{%
\partial y^{i}}}}\right) {w}^{j}${.}

The condition that $\bar{S}$ is adapted to the tangent form $\omega $ reads 
\begin{equation}
h_{ij}S^{j}+f_{i}(t,x^{i},y^{i},z^{i})=0.  \label{eqcompat}
\end{equation}

Notice that the third order semi-spray $S$ gives a system of third order
equations, having the form 
\begin{equation*}
\frac{d^{3}x^{i}}{dt^{3}}+S^{i}(t,x^{i},\frac{dx^{i}}{dt},\frac{d^{2}x^{i}}{%
dt^{2}})=0;
\end{equation*}%
its solutions are the integral curves of the vector field $S$.

We say that the tangent form $\omega $ is \emph{regular} if the local
matrices $(h_{ij}=\frac{\partial \bar{\omega}_{i}}{\partial y^{j}}-\frac{%
\partial \bar{\omega}_{j}}{\partial y^{i}})$ are non-singular. It is easy to
see that this property is free of coordinates.

\begin{pr}
\label{PrPf-regular}If the tangent form $\omega $ is regular, then the
solutions of the generalized Euler-Lagrange equation $\mathcal{E}=0$, where $%
\mathcal{E}$ is given by (\ref{eqdefEL}) are the same solutions of a second
order equation given by a global second order semi-spray $S:I\!\!R\times
T^{2}M\rightarrow T^{3}M$.
\end{pr}

\emph{Proof.} If $\omega $ is regular, then the matrix $(h_{ij})$ is
invertible and let $(h^{ij})=\left( h_{ij}\right) ^{-1}$. The equation (\ref%
{eqcompat}) gives uniquely $S^{j}=h^{ji}f_{i}(t,x^{i},y^{i},z^{i})\ $and a
map $S:$ $I\!\!R\times T^{2}M\rightarrow T^{3}M$, $%
S(t,x^{i},y^{i},z^{i})=(t,x^{i},y^{i},z^{i},S^{i})$, that is well defined
and gives the second order semi-spray $S$. $\Box $

\section{Non-degenerated and regular tangent forms}

We recall that a (first order) tangent form $\omega $ given by (\ref{eqlocom}%
) is

\begin{description}
\item regular if the local matrix $\left( h_{ij}=\frac{\partial \bar{\omega}%
_{i}}{\partial y^{j}}-\frac{\partial \bar{\omega}_{j}}{\partial y^{i}}%
\right) $ is non-singular and

\item non-degenerated if the local matrix $\left( \frac{\partial \bar{\omega}%
_{i}}{\partial y^{j}}\right) $ is non-singular.
\end{description}

The two above regularity conditions are free of coordinates and can be
related as follows.

If $m=\dim M$ is odd, then there are not regular tangent form on $M$, since
a skew symmetric matrix in singular in this case; but there are
non-degenerated tangent forms. For example, let $g$ be a (pseudo-)
Riemannian metric on $M$, $F(x,y)=\frac{1}{2}g_{x}(y,y)$ its energy map and $%
\bar{\omega}_{i}=\frac{\partial F}{\partial y^{i}}$. Then any tangent form
that has $\bar{\omega}=\bar{\omega}_{i}dx^{i}$ a top tangent form is
non-degenerated. Even in even dimension, this top tangent form is
non-degenerated, but never regular, since $\frac{\partial \bar{\omega}_{i}}{%
\partial y^{j}}-\frac{\partial \bar{\omega}_{j}}{\partial y^{i}}=0$.

In even dimensions, there are regular tangent forms that are degenerated.
For example, in $I\!\!R^{2}$, the top tangent form $\bar{\omega}%
(x,y,X,Y)=(X+Y)dx-(X+Y)dy$ is regular, but degenerated.

We say that a tangent $\omega $ form is \emph{biregular} if it is
hyper-non-degenerated and also regular.

Let us extend the definition of a non-degenerated tangent form to higher
order tangent forms and use it in the study of the Euler-Lagrange equation.

For $k\geq 1$, let us denote $T^{k\ast }M=T^{\ast }M\times _{M}T^{k-1}M$,
the fibered product over the base $M$. A $k$-order top tangent form $\bar{%
\omega}:I\!\!R\times T^{k}M\rightarrow I\!\!R\times T^{\ast }M$ gives rise
to a bundle map $\mathcal{L}_{\bar{\omega}}:I\!\!R\times T^{k}M\rightarrow
I\!\!R\times T^{k\ast }M$, $\mathcal{L}_{\bar{\omega}}(t,x^{(k)})=(t,\pi
_{k}(x^{(k)},\bar{\omega}(t,x^{(k)})))$, that we call the \emph{Legendre map}
of $\bar{\omega}$. The \emph{Legendre map} $\mathcal{L}_{\omega }$ of a $k$%
-order tangent form $\omega $ is, by definition, the Legendre map of its
associated top form. The condition that $\mathcal{L}_{\bar{\omega}}$ be a
local diffeomorphism can be read as $\bar{\omega}$ be a \emph{non-degenerated%
} top tangent form. We say that $\bar{\omega}$ is \emph{hyper-non-degenerated%
} if $\mathcal{L}_{\bar{\omega}}$ is a global diffeomorphism. The same
definitions (\emph{Legendre map}, \emph{non-degenerated}, \emph{%
hyper-non-degenerated }and \emph{biregular}) on a tangent form $\omega $ are
the same as referring to its top form $\bar{\omega}$, as above.

We recall that a $k$-order semi-spray on $M$ is a section $S:I\!\!R\times
T^{k}M\rightarrow I\!\!R\times T^{k+1}M$ of the affine bundle $I\!\!R\times
T^{k+1}M\overset{\pi _{k}}{\rightarrow }{}I\!\!R\times T^{k}M$. It can be
regarded as well as a (time dependent) vector field $\Gamma _{0}$ on the
manifold $T^{k}M$, since $T^{k+1}M\subset TT^{k}M$.

Let $\pi _{E}:E\rightarrow M$ be a fibered manifold. For $k\geq 1$, we say
that a \emph{controlled semi-spray of degree }$k$ on $M$ over $E$ is a map $%
\bar{S}:I\!\!R\times E\times _{M}T^{k}M\rightarrow T^{k+1}M$ such that
denoting $\pi ^{\prime }:I\!\!R\times E\times _{M}T^{k}M\rightarrow T^{k}M$
the canonical projection, then $\pi _{k}^{k+1}\circ \bar{S}=\pi ^{\prime }$.
The local coordinate condition for a controlled semi-spray $\bar{S}$ is $\pi
_{k}\circ \bar{S}(t,e,x^{(k)})=x^{(k)}$. We use here only the particular
case of a controlled semi-spray of order $k$ over $E=T^{\ast }M$. We denote $%
T^{k\ast }M=T^{\ast }M\times _{M}T^{k-1}M$, considered as a fibered manifold
over $M$ and we say that a controlled semi-spray $\bar{S}:I\!\!R\times
T^{k\ast }M\rightarrow T^{k}M$ is a $(k-1)$\emph{-order cotangent semi-spray}%
, or a $(k-1)$\emph{-co-semi-spray,} in short.

If a top tangent form of order $k$ is hyper-non-degenerated, then the
inverse $\mathcal{L}_{\bar{\omega}}^{-1}:I\!\!R\times T^{k\ast }M\rightarrow
I\!\!R\times T^{k}M$ of the Legendre map $\mathcal{L}_{\bar{\omega}}$
defines a $(k-1)$-co-semi-spray $\bar{S}:I\!\!R\times T^{k\ast }M\rightarrow
T^{k}M$. For example, if a (first order) top tangent form, or tangent form,
is hyper-non-degenerated, then the inverse of its Legendre map gives a $1$%
-semi-spray $\bar{S}:I\!\!R\times T^{\ast }M\rightarrow TM$.

\subsection{The dynamics of regular and biregular tangent forms}

We prove in this subsection that the dynamics on $M$ of a regular tangent
form comes from the projection of the integral curves of a vector field $X$
on $T^{2\ast }M=T^{\ast }M\times _{M}TM$, while for a biregular tangent
form, its dynamics comes from the projection of the integral curves of a
vector field $Y$ on $T_{2}^{0}M=TM\times _{M}TM$.

A regular (first order) tangent form $\omega $ gives rise to a $2$%
-co-semi-spray $\bar{S}:I\!\!R\times T^{2\ast }M\rightarrow T^{2}M$ as
follows. Let us consider the Ostrogradski tangent form $\Omega =(\frac{%
\partial L_{\omega }}{\partial y^{i}}-\omega _{i})dx^{i}+\frac{\partial
L_{\omega }}{\partial z^{i}}dx^{i}$, $L_{\omega }=\omega _{0}+\omega
_{i}y^{i}+\bar{\omega}_{i}z^{i}$, $\Phi $ be the second order tangent form
given by Proposition \ref{pr-ostr} and $\Phi =\Phi _{i}dx^{i}+\bar{\Phi}%
_{i}dy^{i}$. We have

$\bar{\Phi}_{i}=\frac{\partial L_{\omega }}{\partial y^{i}}-\omega _{i}-%
\frac{d}{dt}\frac{\partial L_{\omega }}{\partial z^{i}}=$ $\frac{\partial
\omega _{0}}{\partial y^{i}}+\frac{\partial \omega _{j}}{\partial y^{i}}%
y^{j}+\frac{\partial \bar{\omega}_{j}}{\partial y^{i}}z^{j}-(\frac{\partial 
\bar{\omega}_{i}}{\partial x^{j}}y^{j}+\frac{\partial \bar{\omega}_{i}}{%
\partial y^{j}}z^{j}+\frac{\partial \bar{\omega}_{i}}{\partial t})=$

$\frac{\partial \omega _{0}}{\partial y^{i}}-\frac{\partial \bar{\omega}_{i}%
}{\partial t}+\left( \frac{\partial \omega _{j}}{\partial y^{i}}-\frac{%
\partial \bar{\omega}_{i}}{\partial x^{j}}\right) y^{j}+\left( \frac{%
\partial \bar{\omega}_{j}}{\partial y^{i}}-\frac{\partial \bar{\omega}_{i}}{%
\partial y^{j}}\right) z^{j}$ and

$\Phi _{i}=\frac{\partial L_{\omega }}{\partial x^{i}}-\frac{d}{dt}\omega
_{i}=\frac{\partial \omega _{0}}{\partial x^{i}}-\frac{\partial \omega i}{%
\partial t}+\left( \frac{\partial \omega _{j}}{\partial x^{i}}-\frac{%
\partial \omega _{i}}{\partial x^{j}}\right) y^{j}+\left( \frac{\partial 
\bar{\omega}_{j}}{\partial x^{i}}-\frac{\partial \omega _{i}}{\partial y^{j}}%
\right) z^{j}$.

Let us consider the map $\mathcal{L}^{\prime }:I\!\!R\times
T^{2}M\rightarrow I\!\!R\times T^{2\ast }M$, $\mathcal{L}^{\prime
}(t,x^{(2)})=$ $(t,$ $\pi _{1}(x^{(2)}),$ $\mathcal{E}_{\Omega
}^{(2)}(x^{(2)}))$, where $\mathcal{E}_{\Omega }^{(2)}:T^{2}M\rightarrow
T^{\ast }M$ is the Lagrange top derivative of $\Omega $.

It is easy to see that $\omega $ is a regular tangent form, i.e. the matrix $%
\left( \frac{\partial \bar{\omega}_{j}}{\partial x^{i}}-\frac{\partial
\omega _{i}}{\partial y^{j}}\right) $ is non-singular, iff $\mathcal{L}%
_{\omega }^{\prime }$ is a global diffeomorphism. Then the inverse of $%
\mathcal{L}_{\omega }^{\prime }$ gives a $2$-co-semi-spray $S:I\!\!R\times
T^{2\ast }M\rightarrow T^{2}M$. The expressions of the functions $S^{i}$
comes from the equations $\bar{\Phi}_{i}=p_{i}$, thus 
\begin{equation}
S^{j}=h^{ij}\left( p_{i}-\frac{\partial \omega _{0}}{\partial y^{i}}+\frac{%
\partial \bar{\omega}_{i}}{\partial t}-\left( \frac{\partial \omega _{j}}{%
\partial y^{i}}-\frac{\partial \bar{\omega}_{i}}{\partial x^{j}}\right)
y^{j}\right) ,  \label{eqsemispr01}
\end{equation}%
where $(h^{ij})=\left( h_{ij}\right) ^{-1}$, $h_{ij}=\frac{\partial \bar{%
\omega}_{j}}{\partial y^{i}}-\frac{\partial \bar{\omega}_{i}}{\partial y^{j}}
$.

Now we can go further, to find the integral curves of the action of $\omega $%
.

Considering local coordinates coming from an open $U\subset M$, then $%
\mathcal{E}_{\omega }^{\prime }=\Omega =\Omega _{i}dx^{i}+\bar{\Omega}%
_{i}dy^{i}$ and let $X_{U}=\frac{\partial }{\partial t}+y^{i}\frac{\partial 
}{\partial x^{i}}+S^{i}(t,x^{j},y^{j},p_{j})\frac{\partial }{\partial y^{i}}+%
\tilde{\Phi}_{i}\frac{\partial }{\partial p_{i}}$ be the expression of a
local vector field on $I\!\!R\times T^{2\ast }M$, where $\tilde{\Phi}%
_{i}(t,x^{i},y^{i},p_{i})=$ $\Phi
_{i}(t,x^{i},y^{i},S^{i}(t,x^{i},y^{i},p_{i}))=$ $\frac{\partial \omega _{0}%
}{\partial x^{i}}-\frac{\partial \omega _{i}}{\partial t}+\left( \frac{%
\partial \omega _{j}}{\partial x^{i}}-\frac{\partial \omega i}{\partial x^{j}%
}\right) y^{j}+\left( \frac{\partial \bar{\omega}_{j}}{\partial x^{i}}-\frac{%
\partial \omega _{i}}{\partial y^{j}}\right) S^{j}$. Let us denote by $\pi
_{\ast 2}:T^{2\ast }M\rightarrow M$ the canonical projection.

\begin{pr}
\label{PrPf-non-deg}Let $\omega $ be a regular (first order) tangent form.
Then:

\begin{enumerate}
\item the local vector fields $X_{U}$ glue together to a vector field $X$ on 
$T^{2\ast }M$ and

\item the integral curves of $X$ projects by $\pi _{\ast 2}$ to all the
critical curves of the action of $\omega $.
\end{enumerate}
\end{pr}

\emph{Proof.} We use local coordinates $(x^{i})$, $(x^{i},y^{i})$, $%
(x^{i},p_{i})$, $(x^{i},y^{i},p_{i})$ and $(x^{i},y^{i},p_{i}$, $X^{i}$, $%
Y^{i}$, $P_{i})$ on $M$, $TM$, $T^{\ast }M$, $T^{2\ast }M=TM\times
_{M}T^{\ast }M$ and $T(TM\times _{M}T^{\ast }M)$ respectively (see the
Appendix). On the intersection domain of two fibered charts, the local
components of $X$ follow next rule: $S^{i^{\prime }}=$ $\frac{\partial
y^{i^{\prime }}}{\partial x^{i}}y^{i}+\frac{\partial x^{i^{\prime }}}{%
\partial x^{i}}S^{i}$ and $\tilde{\Phi}_{i}=\frac{\partial y^{i^{\prime }}}{%
\partial x^{i}}p_{i^{\prime }}+\frac{\partial x^{i^{\prime }}}{\partial x^{i}%
}\tilde{\Phi}_{i^{\prime }}$ respectively. The first rule follows from the
fact that $(S^{i})$ are the components of a $1$-co-semi-spray. The second
rule follows using similar relations \ref{eqchangeoom} for $\Phi _{i}$ and $%
\bar{\Phi}_{i}$: in the second relation we have that $\bar{\Phi}_{i^{\prime
}}=p_{i^{\prime }}$. Thus 1. follows.

Along an integral curve of $X$ we have $\frac{dx^{i}}{dt}=y^{i}$, $\frac{%
dy^{i}}{dt}=S^{i}$ and $\frac{dp_{i}}{dt}=\tilde{\Phi}_{i}$. Since $p_{i}=%
\bar{\Phi}_{i}(t,x^{j},y^{j},S^{j})$ and $\bar{\Phi}_{i}=\frac{\partial L}{%
\partial y^{i}}-\omega _{i}-\frac{d}{dt}\bar{\omega}_{i}$, thus $\frac{dp_{i}%
}{dt}=\frac{d}{dt}(\frac{\partial L}{\partial y^{i}}-\omega _{i}-\frac{d}{dt}%
\bar{\omega}_{i})$, it follows that $\tilde{\Phi}_{i}=$ $\frac{\partial L}{%
\partial x^{i}}-\frac{d}{dt}\omega _{i}=$ $\frac{d}{dt}(\frac{\partial L}{%
\partial y^{i}}-\omega _{i}-\frac{d}{dt}\bar{\omega}_{i})$, i.e. the
Euler-Lagrange equation holds along any curve $t\rightarrow x(t)$. This
proves 2. $\Box $

Let us consider that $\omega $ is biregular, i.e. hyper-non-degenerated and
regular. Let us denote $T_{2}^{0}M=T^{\ast }M\times _{M}T^{\ast }M$ and $\pi
_{02\ast }:T_{2}^{0}M\rightarrow M$ the canonical projection and consider
coordinates $(x^{i},p_{(0)i},p_{(1)i})$ on $T_{2}^{0}M$, induced by the
local coordinates $(x^{i})$ on $M$. We define a map $\mathcal{L}_{\omega
}^{\prime \prime }:I\!\!R\times T^{2}M\rightarrow I\!\!R\times T_{2}^{0}M$, $%
\mathcal{L}_{\omega }^{\prime \prime }(t,x^{(2)})=(t,\bar{\Omega},\bar{\Phi}%
) $, where $\bar{\Omega}=\bar{\omega}$ and $\bar{\Phi}$ are the
corresponding top tangent forms.

Then $\omega $ is non-degenerated and regular iff the map $\mathcal{L}%
_{\omega }^{\prime \prime }$ is a local diffeomorphism. The tangent form $%
\omega $ is hyper-non-degenerated iff $\mathcal{L}_{\omega }^{\prime \prime
} $ is a diffeomorphism

Considering local coordinates as previously, the Legendre map $\mathcal{L}%
_{\omega }^{\prime \prime }$ has the local expression $\mathcal{L}_{\omega
}^{\prime \prime }(t,x^{i},y^{i},z^{i})=(t,\bar{\omega}_{i}(t,x^{i},y^{i}),%
\bar{\Omega}_{i}(t,x^{i},y^{i},z^{i}))$. As above it follows that $\mathcal{L%
}_{\omega }^{\prime \prime }$ is a local diffeomorphism that is a global one
iff the Legendre map is a global diffeomorphism.

If the tangent form $\omega $ is biregular, then considering some local
coordinates $(t,p_{(0)i},p_{(1)i})$ on $I\!\!R\times T_{2}^{0}M$, coming
from an open $U\subset M$, the equations $\bar{\omega}%
_{i}(t,x^{i},y^{i})=p_{(0)i}$ give $y^{i}=T^{i}(t,x^{j},$ $p_{(0)j})$ and
the equations $\bar{\Omega}_{i}(t,x^{i},T^{i},z^{i})=p_{(1)i}$ give $%
z^{i}=S^{i}(t,x^{j},p_{(0)j},p_{(1)j})$. If $\omega $ is hyperregular and
hyper-non-degenerated, then the local functions $(T^{i})$ and $(S^{i})$ come
from some co-semi-sprays and give some global diffeomorphisms $%
T:I\!\!R\times T^{\ast }M\rightarrow I\!\!R\times TM$ and $S:I\!\!R\times
T^{2\ast }M\rightarrow I\!\!R\times T^{2}M$ respectively. Let us consider
the vector field on $T_{2}^{0}U$, given by $Y_{U}=T^{i}\frac{\partial }{%
\partial x^{i}}+(\hat{\Omega}_{i}-p_{(1)i})\frac{\partial }{\partial p_{(0)i}%
}+\hat{\Phi}_{i}\frac{\partial }{\partial p_{(1)i}}$, where $\hat{\Omega}%
_{i}(t,x^{j},p_{(0)j})=\Omega _{i}(t,x^{j},T^{j})$, $\hat{\Phi}%
_{i}(t,x^{j},p_{(0)j},p_{(1)j})=\Phi _{i}(t,x^{j},T^{j},$ $%
S^{j}(t,x^{j},T^{j},p_{(1)j}))$.

\begin{pr}
\label{PrPf-non-deg1}Let $\omega $ be a biregular (first order) tangent
form. Then:

\begin{enumerate}
\item the local vector fields $Y_{U}$ glue together to a vector field $Y$ on 
$T_{2}^{0}M$ and

\item the integral curves of $Y$ projects by $\pi _{\ast 2}$ to all the
critical curves of the action of $\omega $.
\end{enumerate}
\end{pr}

\emph{Proof.} We use local coordinates $(x^{i})$, $(x^{i},p_{(0)i},p_{(1)i})$
and $(x^{i},p_{(0)i},p_{(1)i},x^{i},$ $P_{(0)i},$ $P_{(1)i})$ on $M$, $%
T_{2}^{0}M=T^{\ast }M\times _{M}T^{\ast }M$ and $T(TM^{\ast }\times
_{M}T^{\ast }M)$ respectively (see the Appendix). Then, on the intersection
of two local bundle charts, the local components of $Y$ follow the rules $%
T^{i^{\prime }}=$ $\frac{\partial x^{i^{\prime }}}{\partial x^{i}}T^{i}$, $(%
\hat{\Omega}_{i}-p_{(1)i})=\frac{\partial x^{i^{\prime }}}{\partial x^{i}}(%
\hat{\Omega}_{i^{\prime }}-p_{(1)i^{\prime }})+\frac{\partial y^{i^{\prime }}%
}{\partial x^{i}}p_{(0)i}$ and $\hat{\Phi}_{i}=\frac{\partial x^{i^{\prime }}%
}{\partial x^{i}}\hat{\Phi}_{i^{\prime }}+\frac{\partial y^{i^{\prime }}}{%
\partial x^{i}}p_{(1)i}$. The first relation follows from the fact that $%
(T^{i})$ are the components of a global map in the fibers of $TM$. The
second relation follows from $p_{(1)i}=\frac{\partial x^{i^{\prime }}}{%
\partial x^{i}}p_{(1)i^{\prime }}$ and $\Omega _{i}=\frac{\partial
x^{i^{\prime }}}{\partial x^{i}}\Omega _{i^{\prime }}+\frac{\partial
y^{i^{\prime }}}{\partial x^{i}}p_{(0)i}$ (see the Appendix and definition
of $\hat{\Omega}$). The third relation follows using the definition of $\hat{%
\Theta}$, the second relation \ref{eqchangeoom} and the fact that $\bar{\Phi}%
_{i^{\prime }}=p_{(1)i^{\prime }}$. In order to prove $2.$, along an
integral curve of $Y$ we have $y^{i}=\frac{dx^{i}}{dt}=T^{i}$, $\frac{%
dp_{(0)i}}{dt}=\hat{\Omega}_{i}-p_{(1)i}$ and $\frac{dp_{(1)i}}{dt}=\hat{%
\Theta}_{i}$. According to the definitions, it is easy to prove 2. $\Box $

A full interpretation of the two vector fields $X$ and $Y$ is given in the
next subsection, where we prove that the two vectors are the Hamiltonian
vector fields of two suitable Hamiltonians.

\subsection{Hamiltonian descriptions of biregular tangent forms\label%
{subsect-Hamilton}}

Important tools in describing the dynamic equations of a Hamiltonian system
are offered by quantization. Following similar ideas used in \cite[Section 2.%
]{LSZ01}, one can use Ostrogradski-Dirac and Fadeev-Jakiw methods, but also
a modified Ostrogradski-Dirac method, according to the possibility to
construct constraints slight different from the canonical ones used in
Ostrogradski theory. The Ostrogradski-Dirac method was also used in \cite%
{CMR} to a quantization of a system derived from a Lagrangian affine in
accelerations, involved in the study of a Reegge-Teitelboim model. We give
below a global form of these results. More specifically, we prove in this
subsection that:

-- if $\omega $ is regular and its essential part is time independent, then
there is a symplectic forms $\Xi ^{\prime }$ on $T^{2\ast }M$ and a
Hamiltonian $H:I\!\!R\times T^{2\ast }M\rightarrow I\!\!R$ such that the
Hamiltonian vector field $X_{H}$ gives by projection the dynamics of $\omega 
$ on $M$;

-- if $\omega $ is biregular and its essential part is time independent,
then there is a symplectic forms $\Xi ^{\prime \prime }$ on $T_{2}^{0}M$ and
a Hamiltonian $H^{\prime }:I\!\!R\times T_{2}^{0}M\rightarrow I\!\!R$ such
that the Hamiltonian vector field $X_{H^{\prime }}$ gives by projection the
dynamics of $\omega $ on $M$.

Let us consider the map $\Phi :I\!\!R\times T^{2\ast }M\rightarrow
I\!\!R\times T^{\ast }TM$, $\Phi
(t,x^{i},y^{i},p_{i})=(t,x^{i},y^{i},p_{i}+\omega _{i},\bar{\omega}_{i})$.
Let us denote by $\Phi _{t}:T^{2\ast }M\rightarrow T^{\ast }TM$ the map $%
\Phi _{t}(p^{(2)})=\Phi (t,p^{(2)})$, where $t\in I\!\!R$ is given, and by $%
\Xi $ the canonical symplectic $2$-form on $T^{\ast }TM$. Then we can
consider the induced $2$-form $\Phi _{t}^{\ast }\Xi $ on $T^{2\ast }M$, that
has the local expression $\Phi _{t}^{\ast }\Xi =dx^{i}\wedge (dp_{i}+d\omega
_{i})+dy^{i}\wedge d\bar{\omega}_{i}$, where the differential $d$ is
considered on $T^{2\ast }M$.

\begin{pr}
\label{prHamReg}Let $\omega $ be a (first order) tangent form. For every $%
t\in I\!\!R$ the form $\Xi _{t}^{\prime }=\Phi _{t}^{\ast }\Xi $ is closed
on $T^{2\ast }M$ and it is non-degenerated iff $\omega $ is a regular
tangent form.
\end{pr}

\emph{Proof.} The form $\Phi _{t}^{\ast }\Xi $ is closed since the form $\Xi 
$ is closed. Using local coordinates as above, we have:

$\Phi _{t}^{\ast }\Xi =dx^{i}\wedge (dp_{i}+\frac{\partial \omega _{i}}{%
\partial x^{j}}dx^{j}+\frac{\partial \omega _{i}}{\partial y^{j}}%
dy^{j})+dy^{i}\wedge (\frac{\partial \bar{\omega}_{i}}{\partial x^{j}}dx^{j}+%
\frac{\partial \bar{\omega}_{i}}{\partial y^{j}}dy^{j})=$ $dx^{i}\wedge
dp_{i}+\frac{\partial \omega _{i}}{\partial x^{j}}dx^{i}\wedge dx^{j}+\left( 
\frac{\partial \omega _{i}}{\partial y^{j}}-\frac{\partial \bar{\omega}_{j}}{%
\partial x^{i}}\right) dx^{i}\wedge dy^{j}+\frac{\partial \bar{\omega}_{i}}{%
\partial y^{j}}dy^{i}\wedge dy^{j}$. Thus, using the local base $%
\{dx^{i}\wedge dx^{j}$, $dx^{i}\wedge dy^{j}$, $dx^{i}\wedge dp_{j}$, $%
dy^{i}\wedge dy^{j}$, $dy^{i}\wedge dp_{j}$, , $dp_{i}\wedge dp_{j}\}_{i<j}$%
, then $\Phi _{t}^{\ast }\Xi $ has the matrix%
\begin{equation*}
\left( 
\begin{array}{ccc}
A & B & I \\ 
-B & C & 0 \\ 
-I & 0 & 0%
\end{array}%
\right)
\end{equation*}%
where $A=\left( \frac{\partial \omega _{i}}{\partial x^{j}}-\frac{\partial
\omega _{j}}{\partial x^{i}}\right) $, $B=\left( \frac{\partial \omega _{i}}{%
\partial y^{j}}-\frac{\partial \bar{\omega}_{j}}{\partial x^{i}}\right) $, $%
I=(\delta _{j}^{i})$, $C=\left( \frac{\partial \bar{\omega}_{i}}{\partial
y^{j}}-\frac{\partial \bar{\omega}_{j}}{\partial y^{i}}\right) $. The above
matrix is non-degenerate iff the matrix $C$ is non-degenerate, i.e. iff $%
\omega $ is a regular tangent form. $\Box $

In the case when the essential part of a tangent form on $M$ is time
independent, we can avoid the use of parameter $t$ and then consider the
induced $2$-form $\Xi ^{\prime }=\Phi ^{\prime \ast }\Xi $ on $T^{2\ast }M$,
where $\Phi (t,p^{(2)})=(t,\Phi ^{\prime }(p^{(2)}))$. As in Proposition \ref%
{prHamReg}, the two form $\Xi ^{\prime }$ is a symplectic form on $T^{2\ast
}M$.~We prove now that $\Xi ^{\prime }$ can be used to quantify the
Hamiltonian system derived from a Lagrangian affine in accelerations that
comes from a non-degenerate tangent form.

\begin{theor}
\label{thquant01}Let $\omega $ be a regular (first order) tangent form on $M$
such that its essential part is time independent. Then there are a
symplectic form $\Xi ^{\prime }$ on $T^{2\ast }M$ and a Hamiltonian $%
H:I\!\!R\times T^{2\ast }M\rightarrow I\!\!R$ such that the Hamiltonian
vector field $X_{H}$ is $X$ from Proposition \ref{PrPf-non-deg}.
\end{theor}

\emph{Proof.} According to Proposition \ref{PrPf-non-deg}, it suffices to
prove that the vector field $X=\frac{\partial }{\partial t}+y^{i}\frac{%
\partial }{\partial x^{i}}+S^{i}(t,x^{i},y^{i},p_{i})\frac{\partial }{%
\partial y^{i}}+\tilde{\Phi}_{i}\frac{\partial }{\partial p_{i}}$ is the
Hamiltonian vector field of a suitable Hamiltonian, namely the Hamiltonian $%
H=-p_{i}y^{i}+\omega _{0}$. It reads that $i_{X}\Xi ^{\prime }=dH$. Indeed,
using local coordinates, we have:

$\left( \frac{\partial \omega _{i}}{\partial x^{j}}-\frac{\partial \omega
_{j}}{\partial x^{i}}\right) y^{j}+\left( \frac{\partial \omega _{i}}{%
\partial y^{j}}-\frac{\partial \bar{\omega}_{j}}{\partial x^{i}}\right)
S^{j}+\delta _{i}^{j}\tilde{\Phi}_{j}=$

$\left( \frac{\partial \omega _{i}}{\partial x^{j}}-\frac{\partial \omega
_{j}}{\partial x^{i}}\right) y^{j}+\left( \frac{\partial \omega _{i}}{%
\partial y^{j}}-\frac{\partial \bar{\omega}_{j}}{\partial x^{i}}\right)
S^{j}+$

$\frac{\partial \omega _{0}}{\partial x^{i}}-\frac{\partial \omega i}{%
\partial t}+\left( \frac{\partial \omega _{j}}{\partial x^{i}}-\frac{%
\partial \omega _{i}}{\partial x^{j}}\right) y^{j}+\left( \frac{\partial 
\bar{\omega}_{j}}{\partial x^{i}}-\frac{\partial \omega _{i}}{\partial y^{j}}%
\right) S^{j}=$

$\frac{\partial \omega _{0}}{\partial x^{i}}-\frac{\partial \omega i}{%
\partial t}=\frac{\partial \omega _{0}}{\partial x^{i}}=\frac{\partial H}{%
\partial x^{i}}$,

$-\left( \frac{\partial \omega _{j}}{\partial y^{i}}-\frac{\partial \bar{%
\omega}_{i}}{\partial x^{j}}\right) y^{j}+\left( \frac{\partial \bar{\omega}%
_{i}}{\partial y^{j}}-\frac{\partial \bar{\omega}_{j}}{\partial y^{i}}%
\right) S^{j}=$

$-\left( \frac{\partial \omega _{j}}{\partial y^{i}}-\frac{\partial \bar{%
\omega}_{i}}{\partial x^{j}}\right) y^{j}-p_{i}+\frac{\partial \omega _{0}}{%
\partial y^{i}}-\frac{\partial \bar{\omega}_{i}}{\partial t}+\left( \frac{%
\partial \omega _{j}}{\partial y^{i}}-\frac{\partial \bar{\omega}_{i}}{%
\partial x^{j}}\right) y^{j}=$

$-p_{i}+\frac{\partial \omega _{0}}{\partial y^{i}}-\frac{\partial \bar{%
\omega}_{i}}{\partial t}=-p_{i}+\frac{\partial \omega _{0}}{\partial y^{i}}=%
\frac{\partial H}{\partial y^{i}}$ and

$-\delta _{j}^{i}$\-$y^{j}=-y^{i}=\frac{\partial H}{\partial p_{i}}$. $\Box $

In the case when the essential part $\tilde{\omega}=\omega _{i}dx^{i}+\bar{%
\omega}_{i}dy$ of $\omega $ is not necessarily time independent, the general
formula reads $i_{X}\Xi =dH-\frac{\partial }{\partial t}\tilde{\omega}$,
where $\frac{\partial }{\partial t}\tilde{\omega}=\frac{\partial \omega i}{%
\partial t}dx^{i}+\frac{\partial \bar{\omega}i}{\partial t}dy^{i}$ is a $1$%
--form on $T^{2\ast }M$ induced by the canonical projection $T^{2\ast
}M\rightarrow TM$, by a $1$--form given by the same formula.

Let $\bar{\omega}=\bar{\omega}_{i}dx^{i}$ be a hyper-non-degenerated (first
order) top tangent form, i.e. the Legendre map $\mathcal{L}_{\bar{\omega}%
}:I\!\!R\times TM\rightarrow I\!\!R\times T^{\ast }M$ is a global
diffeomorphism. Then $\mathcal{L}_{\bar{\omega}}^{-1}:I\!\!R\times T^{\ast
}M\rightarrow I\!\!R\times TM$ has the form $(t,x^{i},p_{i})\overset{%
\mathcal{L}_{\bar{\omega}}^{-1}}{\rightarrow }%
{}(t,x^{i},T^{i}(t,x^{i},p_{i}))$. Considering the non-degenerated matrices $%
\left( h_{ij}=\frac{\partial \bar{\omega}_{i}}{\partial y^{j}}\right) $ and
its inverse $\left( \bar{h}^{ij}=\frac{\partial T^{i}}{\partial p_{j}}%
\right) $, we say that $\bar{\omega}$ is co-regular if the matrix $\left( 
\tilde{h}^{ij}=\frac{\partial T^{i}}{\partial p_{j}}-\frac{\partial T^{j}}{%
\partial p_{i}}\right) $ is non-singular in every point of $I\!\!R\times
T^{\ast }M$. We recall that $\bar{\omega}$ is regular if the matrix 
\begin{equation*}
\left( \tilde{h}_{ij}=\frac{\partial \bar{\omega}_{i}}{\partial y^{j}}-\frac{%
\partial \bar{\omega}_{j}}{\partial y^{i}}\right)
\end{equation*}%
is non-singular in every point of $I\!\!R\times TM$.

We say that a tangent form $\omega $ is co-regular if its top tangent form $%
\bar{\omega}$ is co-regular.

\begin{pr}
If a tangent form $\omega $ is non-degenerated then $\omega $ is co-regular
iff $\omega $ is regular.
\end{pr}

\emph{Proof.} Denoting $H=\left( h_{ij}\right) $, $\bar{H}=\left( \bar{h}%
^{ij}\right) =H^{-1}$, then $\left( \tilde{h}^{ij}\right) =\bar{H}-\bar{H}%
^{t}=$ $\bar{H}(H^{t}-H)\bar{H}^{t}$, thus $\left( \tilde{h}^{ij}\right) =%
\bar{H}-\bar{H}^{t}$ is invertible iff $\left( h_{ij}-h_{ji}\right) =H-H^{t}$
is invertible; this prove the assertion. $\Box $

Let us suppose that the tangent form $\omega $ is biregular, i.e.
hyper-non-degenerated and regular. Thus there are some global co-semi-sprays
that give some global diffeomorphisms $T:I\!\!R\times T^{\ast }M\rightarrow
I\!\!R\times TM$ and $S:I\!\!R\times T^{2\ast }M\rightarrow I\!\!R\times
T^{2}M$ respectively. We consider the local functions $(T^{i})$ and $(S^{i})$
that come from these co-semi-sprays.

Let us consider the diffeomorphism $\Psi :I\!\!R\times T_{2}^{0}M\rightarrow
I\!\!R\times T^{\ast }TM$, 
\begin{equation*}
\Psi (t,x^{i},p_{(0)i},p_{(1)i})=(t,x^{i},T^{i}(t,x^{j},p_{(0)j}),p_{(1)i}+%
\tilde{\omega}_{i}(t,x^{j},p_{(0)j}),p_{(0)i}),
\end{equation*}%
where $\tilde{\omega}_{i}(t,x^{j},p_{(0)j})=\omega
_{i}(t,x^{j},T^{j}(t,x^{j},p_{(0)j}))$. Let us denote by $\Psi
_{t}:T_{2}^{0}M\rightarrow T^{\ast }TM$ the map $\Psi
_{t}(x^{i},p_{(0)i},p_{(1)i})=\Psi (t,x^{i},p_{(0)i},p_{(1)i})$, where $t\in
I\!\!R$ is given, and by $\Xi $ the canonical symplectic $2$-form on $%
T^{\ast }TM$. Then we can consider the induced $2$-form $\Psi _{t}^{\ast
}\Xi $ on $T_{2}^{0}M$, that has the local expression $\Phi _{t}^{\ast }\Xi
=dx^{i}\wedge (p_{(1)i}+d\tilde{\omega}_{i})+dT^{i}\wedge dp_{(0)i}$, where
the differential $d$ is considered on $T_{2}^{0}M$.

Let us denote by $\digamma :I\!\!R\times T_{2}^{0}M\rightarrow I\!\!R\times
T^{2\ast }M$ the diffeomorphism given by $\digamma
(t,x^{i},p_{(0)i},p_{(1)i})=(t,x^{i},T^{i}(t,x^{j},p_{(0)j}),p_{(1)i})$,
provided that $\omega $ is hyper-non-degenerated. It is easy to see that $%
\Psi _{t}=\Phi _{t}\circ \digamma _{t}$, thus $\Psi _{t}^{\ast }\Xi
=\digamma _{t}^{\ast }\Phi _{t}^{\ast }\Xi =$ $\digamma _{t}^{\ast }\Xi
_{t}^{\prime }$. In a similar way as Proposition \ref{prHamReg}, the
following statement holds true.

\begin{pr}
\label{prsym02}Let $\omega $ be a biregular tangent form. For every $t\in
I\!\!R$ the two form $\Xi _{t}^{\prime \prime }=\Psi _{t}^{\ast }\Xi
=\digamma _{t}^{\ast }\Xi _{t}^{\prime }$ is a symplectic form on $%
T_{2}^{0}M $.
\end{pr}

Using local coordinates as above, we have:

$\Psi _{t}^{\ast }\Xi =dx^{i}\wedge (dp_{(1)i}+\frac{\partial \tilde{\omega}%
_{i}}{\partial x^{j}}dx^{j}+\frac{\partial \tilde{\omega}_{i}}{\partial
p_{(0)j}}dp_{(0)j})+(\frac{\partial T^{i}}{\partial x^{j}}dx^{j}+\frac{%
\partial T^{i}}{\partial p_{(0)j}}dp_{(0)j})\wedge dp_{(0)i}=$

$dx^{i}\wedge dp_{(1)i}+\frac{\partial \tilde{\omega}_{i}}{\partial x^{j}}%
dx^{i}\wedge dx^{j}+\left( \frac{\partial \tilde{\omega}_{i}}{\partial
p_{(0)j}}-\frac{\partial T^{j}}{\partial x^{i}}\right) dx^{i}\wedge
dp_{(0)j}+\frac{\partial T^{j}}{\partial p_{(0)i}}dp_{(0)i}\wedge dp_{(0)j}$%
. Thus, using the local base $\{dx^{i}\wedge dx^{j}$, $dx^{i}\wedge
dp_{(0)j} $, $dx^{i}\wedge dp_{(1)j}$, $dp_{(0)i}\wedge dp_{(0)j}$, $%
dp_{(0)i}\wedge dp_{(1)j}$, , $dp_{(1)i}\wedge dp_{(1)j}\}_{i<j}$, then $%
\Phi _{t}^{\ast }\Xi $ has the matrix%
\begin{equation*}
\left( 
\begin{array}{ccc}
A^{\prime } & B^{\prime } & I \\ 
-B^{\prime } & C^{\prime } & 0 \\ 
-I & 0 & 0%
\end{array}%
\right) ,
\end{equation*}%
where $A^{\prime }=\left( \frac{\partial \tilde{\omega}_{i}}{\partial x^{j}}-%
\frac{\partial \tilde{\omega}_{j}}{\partial x^{i}}\right) $, $B^{\prime
}=\left( \frac{\partial \tilde{\omega}_{i}}{\partial p_{(0)j}}-\frac{%
\partial T^{j}}{\partial x^{i}}\right) $, $C^{\prime }=\left( \frac{\partial
T^{j}}{\partial p_{(0)i}}-\frac{\partial T^{i}}{\partial p_{(0)j}}\right) $
and $I=(\delta _{ij})$. The above matrix is non-degenerated iff the matrix $%
C^{\prime }$ is non-degenerate i.e. iff $\omega $ is a biregular tangent
form.

In the case when the essential part of a tangent form on $M$ is
hyper-non-degenerated and time independent, we can avoid the use of
parameter $t$ and then consider the diffeomorphisms $\Psi ^{\prime
}:T_{2}^{0}M\rightarrow T^{\ast }TM$ and $\digamma ^{\prime
}:T_{2}^{0}M\rightarrow T^{2\ast }M$, induced by $\Psi $ and $\digamma $, as
in the previous case of $\Phi $ induced by $\Phi ^{\prime }$. We have $\Psi
=\Phi \circ \digamma $, thus $\Psi ^{\prime \ast }\Xi =\digamma ^{\prime
\ast }\Phi ^{\prime \ast }\Xi =$ $\digamma ^{\prime \ast }\Xi ^{\prime }$.
Notice that $\Psi (t,x^{i},p_{(0)i},p_{(1)i})=(t,\Psi ^{\prime
}(x^{i},p_{(0)i},p_{(1)i}))$ $\Phi (t,p^{(2)})=(t,\Phi ^{\prime }(p^{(2)}))$%
. As in Proposition \ref{prsym02}, the two form $\Xi ^{\prime \prime }=\Psi
^{\prime \ast }\Xi =\digamma ^{\prime \ast }\Xi ^{\prime }$ is a symplectic
form on $T_{2}^{0}M$. We prove now that $\Xi ^{\prime \prime }$ can be used
also to quantify the Hamiltonian system derived from a Lagrangian affine in
accelerations that comes from a non-degenerate tangent form. Using Theorem %
\ref{thquant01}, we can prove the following statement.

\begin{theor}
\label{thquant02}Let $\omega $ be a biregular (first order) tangent form on $%
M$ such that its essential part is time independent. Then there are a
symplectic form $\Xi ^{\prime \prime }$ on $T_{2}^{0}M$ and a Hamiltonian $%
H^{\prime }:I\!\!R\times T_{2}^{0}M\rightarrow I\!\!R$ such that the
Hamiltonian vector field $X_{H^{\prime }}$ is $Y$ from Proposition \ref%
{PrPf-non-deg1}.
\end{theor}

\emph{Proof.} It suffices to prove that the vector fields $X=X_{H}\in 
\mathcal{X}(T^{2\ast }M)$ used in Theorem \ref{thquant01} and $Y\in \mathcal{%
X}(T_{2}^{0}M)$ used in Proposition \ref{PrPf-non-deg1} are related by the
diffeomorphisms $\digamma ^{\prime }:T_{2}^{0}M\rightarrow T^{2\ast }M$ ,
i.e. $(\digamma ^{\prime })_{\ast }Y\circ (\digamma ^{\prime -1})=X$, or $%
(\digamma ^{\prime -1})_{\ast }X\circ (\digamma ^{\prime })=Y$.

Indeed, using local coordinates, $(\digamma ^{\prime -1})_{\ast }$ has the
local matrix%
\begin{equation*}
\left( 
\begin{array}{ccc}
I & 0 & 0 \\ 
D & E & 0 \\ 
0 & 0 & I%
\end{array}%
\right) ,
\end{equation*}%
where $I=\left( \delta _{j}^{i}\right) $, $D=\left( \frac{\partial \bar{%
\omega}_{i}}{\partial x^{j}}\right) $, $E=\left( \frac{\partial \bar{\omega}%
_{i}}{\partial y^{j}}\right) $. Then $(\digamma ^{\prime -1})_{\ast }X$ and $%
Y^{\prime }=(\digamma ^{\prime -1})_{\ast }X\circ (\digamma ^{\prime })$
have the local expressions

$(\digamma ^{\prime -1})_{\ast }X=y^{i}\frac{\partial }{\partial x^{i}}%
+(y^{j}\frac{\partial \bar{\omega}_{i}}{\partial x^{j}}+S^{j}\frac{\partial 
\bar{\omega}_{i}}{\partial y^{j}})\frac{\partial }{\partial p_{(0)i}}+\bar{%
\Phi}_{i}\frac{\partial }{\partial p_{(1)i}}$ and

\noindent$Y^{\prime }=T^{i}\frac{\partial }{\partial x^{i}}+(T^{i}\frac{%
\partial \bar{\omega}_{i}}{\partial x^{j}}%
(t,x^{i},T^{i})+S^{j}(t,x^{j},T^{j},p_{(1)j})\frac{\partial \bar{\omega}_{i}%
}{\partial y^{j}}(t,x^{i},T^{i}))\frac{\partial }{\partial p_{(0)i}}+\hat{%
\Phi}_{i}\frac{\partial }{\partial p_{(1)i}}=$

$T^{i}\frac{\partial }{\partial x^{i}}+(\hat{\Omega}_{i}-p_{(1)i})\frac{%
\partial }{\partial p_{(0)i}}+\hat{\Phi}_{i}\frac{\partial }{\partial
p_{(1)i}}$, since

$y^{i}\frac{\partial \bar{\omega}_{i}}{\partial x^{j}}+S^{j}\frac{\partial 
\bar{\omega}_{i}}{\partial y^{j}}=\frac{\partial L:}{\partial y^{i}}-\omega
_{i}-p_{i}=\Omega _{i}-p_{i}$.

Thus $Y^{\prime }=Y\in \mathcal{X}(T_{2}^{0}M)$, used in Proposition \ref%
{PrPf-non-deg1}.

Notice that the pull-back of the Hamiltonians $H=-p_{i}y^{i}+\omega _{0}$ by 
$\digamma _{t}$ is the Hamiltonian $H^{\prime }:I\!\!R\times
T_{2}^{0}M\rightarrow I\!\!R$, $H^{\prime
}(t,x^{i},p_{(0)i},p_{(1)i})=-p_{(1)i}T^{i}(x^{j},p_{(0)j})+\omega
_{0}(t,x^{i},T^{i})$. $\Box $

\section{Some examples and special cases\label{subsectEx}}

We say that a tangent form $\omega \in \mathcal{X}^{\ast }(I\!\!R\times TM)$
is \emph{singular} if it is locally equivalent to a local Lagrangian.

\begin{pr}
A tangent form is \emph{singular }iff its top component $\bar{\omega}$,
viewed as a vertical form, is vertical closed.
\end{pr}

\emph{Proof.} The tangent form $\omega =\omega _{0}dt+\omega _{i}dx^{i}+\bar{%
\omega}_{i}dy^{i}=$ $(\omega _{0}+y^{i}\omega _{i})dt+\omega _{i}\delta
x^{i}+\bar{\omega}_{i}dy^{i}$ is locally equivalent to a local Lagrangian
form iff locally its top component $\bar{\omega}_{i}$ has the form $\bar{%
\omega}_{i}=\frac{\partial \mu }{\partial y^{i}}$. Using Poincar\'{e} Lemma,
this condition is equivalent to ${{\frac{\partial \bar{\omega}_{i}}{\partial
y^{j}}-\frac{\partial \bar{\omega}_{j}}{\partial y^{i}}=0}}$, i.e. $\bar{%
\omega}=\bar{\omega}_{i}dx^{i}$ is vertically closed. $\Box $

We say that $\omega $ is:

\begin{description}
\item \emph{globally singular }if there are two Lagrangians $L_{0}$, $%
L_{1}:I\!\!R\times TM\rightarrow I\!\!R$ and a top Lagrangian form $\mu =\mu
_{i}dx^{i}$, such that $\omega -L_{0}dt=\mu _{i}\delta x^{i}+dL_{1}$;

\item \emph{locally singular }if there is a Lagrangian $L_{0}:I\!\!R\times
TM\rightarrow I\!\!R$, a closed form $\omega _{0}\in \mathcal{X}^{\ast
}(I\!\!R\times TM)$ and a top Lagrangian form $\mu =\mu _{i}dx^{i}$, such
that $\omega -L_{0}dt=\mu _{i}\delta x^{i}+\omega _{0}$.
\end{description}

It is easy to see that if the tangent form $\omega $ is globally or locally
singular it is also singular.

A (\emph{global) non-Lagrangian system} is given by a tangent form $\omega $
for which there are two Lagrangians $L,\mu _{0}:TM\rightarrow I\!\!R$ and a
top tangent form $\mu =\mu _{i}dx^{i}$, such that $\omega -dL=\mu _{0}dt+\mu
_{i}dx^{i}$, thus $\omega _{0}={{\frac{\partial L}{\partial t}}}+\mu _{0}$, $%
\omega _{i}={{\frac{\partial L}{\partial x^{i}}}}+\mu _{i}$ and $\bar{\omega}%
_{i}={{\frac{\partial L}{\partial y^{i}}}}$. Since $\omega -(\mu
_{0}+y^{i}\mu _{i})dt=$ $\mu _{i}\delta x^{i}+dL$, it follows that a
(global) non-Lagrangian system is equivalent to give a global singular
tangent form.

We can relax the above condition defining a \emph{local non-Lagrangian system%
} as a tangent form $\omega $ such that $\omega -\tilde{\omega}=\mu
_{0}dt+\mu _{i}dx^{i}$, where $\tilde{\omega}$ is a closed form and $\mu $, $%
\mu _{0}\ $are as previously. In the same way, it follows that a local
non-Lagrangian system is equivalent to give a locally singular tangent form.

If $\omega $ is a local non-Lagrangian system on $TM$, then it can be proved
that it is a global one.

In the case when $\omega $ is differentiable only on $TM_{\ast
}=TM\backslash \{\bar{0}\}$, where $\{\bar{0}\}$ is the image of the null
section, then it makes sense to mark the difference between a local and a
global tangent form.

For example, the tangent form $\omega =\frac{XY}{\sqrt{\left(
X^{2}+Y^{2}\right) ^{3}}}dx-\frac{X^{2}}{\sqrt{\left( X^{2}+Y^{2}\right) ^{3}%
}}dy$ is a local non-Lagrangian on $I\!\!R^{2}\times I\!\!R_{\ast }^{2}$,
but not a global one.

Instead of $TM_{\ast }$ one can consider another open submanifold of $TM$.

An other example: the tangent form $\omega =Ldt$, associated with a
non-constant Lagrangian $L$, defines a non-Lagrangian system.

Some important classes of tangent forms are:

-- When $\omega_{0}=0$; for example, this is the case of time independent
Lagrangians $L=L(x^{i},y^{i})$, since $\omega_{0}={{{\ {\frac{\partial L }{%
\partial t}}}}}$;

-- When $\omega _{0}=\omega _{i}=0$; for example, this is the case of
Lagrangians that depend only on direction: $L=L(y^{i})$.

If $\omega =\bar{\omega}_{j}(y^{i})dy$\-$^{j}$, then the equation (\ref%
{EL-ord-2+}) has the form ${{{{\frac{d}{dt}(}}\frac{\partial \bar{\omega}_{j}%
}{\partial y^{i}}{\frac{d^{2}x_{0}^{j}}{dt^{2}}}}})-{{{{\frac{d^{2}}{dt^{2}}}%
}}}\bar{\omega}_{i}=0$, or ${{\frac{\partial \bar{\omega}_{j}}{\partial y^{i}%
}{\frac{d^{2}x_{0}^{j}}{dt^{2}}}}}-{{{{\frac{d}{dt}}}}}\bar{\omega}%
_{i}=c_{i} $ $\Leftrightarrow $ $({{\frac{\partial \bar{\omega}_{j}}{%
\partial y^{i}}}}-{{\frac{\partial \bar{\omega}_{j}}{\partial y^{i}}){\frac{%
d^{2}x_{0}^{j}}{dt^{2}}}}}=c_{i}$.

\textbf{Example 1.} Let us consider coordinates $(x,y)$ on $I\!\!R^{2}$ and $%
(x,y,X,Y)$ on $I\!\!R^{4}=TI\!\!R^{2}$. Let $\omega =YdX-XdY$. The equations
(\ref{EL-ord-2+}) have the form: $-{{\frac{d}{dt}}}\left( {{\frac{d^{2}y}{%
dt^{2}}}}\right) -{{\frac{d^{2}}{dt^{2}}}}\left( {{\frac{dy}{dt}}}\right) =0$%
, or ${{{\frac{d^{3}y}{d^{3}t}}=0}}$, and ${{\frac{d}{dt}}}\left( {{\frac{%
d^{2}x}{dt^{2}}}}\right) +{{\frac{d^{2}}{dt^{2}}}}\left( {{\frac{dx}{dt}}}%
\right) =0$, or ${{{\frac{d^{3}x}{d^{3}t}}=0}}$. The exact solution is: $%
x(t)=C_{1}+C_{2}t+C_{3}t^{2}$, $y(t)=C_{4}+C_{5}t+C_{6}t^{2}$.

\textbf{Example 2.} In $I\!\!R^{2}$, as in Example 1. above, let $%
\omega=-ydx+xdy+YdX-XdY$. The equations (\ref{EL-ord-2+}) have the form ${{%
\frac{\partial\omega_{j}}{\partial x^{i}}\frac{dx_{0}^{j}}{dt}}}-{{{{\frac{d%
}{dt}(}}}}\omega_{i}+{{\frac{\partial\bar{\omega}_{j}}{\partial y^{i}}\frac{%
d^{2}x_{0}^{j}}{dt^{2}})}}+{{{{\frac{d^{2}}{dt^{2}}}}}}\bar {\omega}_{i}=0.$

For $j=1$, ${{\frac{dy}{dt}-\frac{d}{dt}(-y-{\frac{d^{2}y}{dt^{2}}})}}+{{%
\frac{d^{2}}{dt^{2}}}}\left( {{\frac{dy}{dt}}}\right) =0$, or ${{\frac{dy}{dt%
}}}+{{{\frac{d^{3}y}{d^{3}t}}=0}}$ and

For $j=2$, $-{{\frac{dx}{dt}-\frac{d}{dt}(x+{\frac{d^{2}x}{dt^{2}}})}}-{{%
\frac{d^{2}}{dt^{2}}}}\left( {{\frac{dx}{dt}}}\right) =0$, or ${{\frac{dx}{dt%
}}}+{{{\frac{d^{3}x}{d^{3}t}}=0}}$. $\allowbreak$

The general solution is $x(t)=c_{1}\cos t+c_{3}\sin t+c_{5}$, $%
x(t)=c_{2}\cos t+c_{4}\sin t+c_{6}$. The integral curves are ellipses and
straight lines. If $t_{1}<t_{2}<t_{3}$ are given, then for every three
distinct points $A_{\alpha }(x_{\alpha },y_{\alpha })\in I\!\!R^{2}$, $%
\alpha =\overline{1,3}$, there is a unique integral curve in the family that
contains the three points, i.e. $t\rightarrow (x(t),y(t))$, $x(t_{\alpha
})=x_{\alpha }$, $y(t_{\alpha })=y_{\alpha }$, $\alpha =\overline{1,3}$.

{This feature characterizes the dynamics generated by a third order
differential equation, when in general, an integral curve is determined by
three distinct points. Let us notice that for a second order differential
equation, an integral curve is determined, in general, by two distinct
points. }

Let us consider now the case $\dim M=1$. In this case, since the only
skew-symmetric matrix of first order is the null matrix, the equation (\ref%
{EL-ord-2+}) is always of second order, for every tangent form $\tilde{\omega%
}=\omega _{0}dt+\omega dx+\bar{\omega}dy$, having the form

${(}\frac{\partial ^{2}\bar{\omega}}{\partial t\partial y}{-2}\frac{\partial
\omega }{\partial y}{{{+}2\frac{\partial \bar{\omega}}{\partial x}}}){{\frac{%
d^{2}x_{0}}{dt^{2}}}}+{{\frac{\partial ^{2}\bar{\omega}}{\partial x^{2}}({{%
\frac{dx_{0}}{dt}}})}}^{2}+$ $(-{{\frac{\partial ^{2}\omega }{\partial
t\partial y}-\frac{\partial ^{2}\omega }{\partial x\partial y}-\frac{%
\partial \omega }{\partial y^{2}}+2\frac{\partial ^{2}\bar{\omega}}{\partial
x\partial t}){\frac{dx_{0}}{dt}}}}+$ ${{\frac{\partial \omega _{0}}{\partial
x}}}-$ ${{\frac{\partial ^{2}\omega _{0}}{\partial t\partial y}}}-$ ${{{{%
\frac{\partial \omega }{\partial t}}}}}+$ $\frac{\partial ^{2}\bar{\omega}}{%
\partial t^{2}}=0$.

In the case when the local functions $\omega _{0}$, $\omega $ and $\bar{%
\omega}$ do not depend on $y$, the above equation becomes%
\begin{equation}
2{{{{\frac{\partial \bar{\omega}}{\partial x}}}\frac{d^{2}x_{0}}{dt^{2}}+{{%
\frac{\partial ^{2}\bar{\omega}}{\partial x^{2}}}}}}\left( {{\frac{dx_{0}}{dt%
}}}\right) ^{2}{{+2\frac{\partial ^{2}\bar{\omega}}{\partial x\partial t}{%
\frac{dx_{0}}{dt}+{\frac{\partial \omega _{0}}{\partial x}-\frac{\partial
\omega }{\partial t}+\frac{\partial ^{2}\bar{\omega}}{\partial t^{2}}=0.}}}}
\label{eqEuLag1}
\end{equation}

We can give a global description of this fact. It well-known that any one
dimensional manifold is diffeomorphic with $I\!\!R$ or $S^{1}$. On $I\!\!R$
one can take a single global chart, while on $S^{1}$ one can take two
charts, where the coordinate functions follow the rule ${\ {\frac{\partial
x^{\prime }}{\partial x}}}=\pm 1$. Using the rules that coordinates follow,
it follows that if $\omega _{0}$, $\omega $ and $\bar{\omega}$ do not depend
on $y$ on the domains of the two local charts, this is true on the
intersection domain; we call a such tangent form $\tilde{\omega}$ as a \emph{%
basic tangent form}. We suppose also that ${{{{\frac{\partial \bar{\omega}}{%
\partial x}\neq 0}}}}$ in every point, thus $\tilde{\omega}$ is regular

According to \cite[Section 2.]{CN}, a \emph{standard} Lagrangian has the
form 
\begin{equation}
L(t,x,y)=\frac{1}{2}P(x,t)y^{2}+Q(x,t)y+R(x,t).  \label{standLag}
\end{equation}
Its Euler-Lagrange equation is $2Px^{\prime\prime}+P_{x}\left( x^{\prime
}\right) ^{2}+2P_{t}x^{\prime}+2(Q_{t}-R_{x})=0$, where subscripts $x$, $t$
denote partial derivatives and $x^{\prime}={{{{\frac{dx}{dt}}}}}$, $%
x^{\prime\prime}={{{{\frac{d^{2}x}{dt^{2}}}}}}$. In \cite[Proposition 2.1.]%
{CN} one prove that a second order equation%
\begin{equation*}
x^{\prime\prime}+a(t,x)\left( x^{\prime}\right) ^{2}+b(t,x)x^{\prime
}+c(t,x)=0
\end{equation*}
admits a standard Lagrangian description (\ref{standLag}) iff $b_{x}=2a_{t}$%
; then $P=$ $\exp(2\int^{x}a(t,$ $s)ds)$ and $R=%
\int^{x}(Q_{t}(t,s)-c(t,s)P(t,s))ds$, where $Q=Q(x,t)$ is an arbitrary
function. The following result can be proved by a straightforward
verification.

\begin{pr}
\label{prM=1}The generalized Euler-Lagrange equation of a regular and basic
tangent form on a one dimensional manifold admits locally standard
Lagrangian descriptions.
\end{pr}

\emph{Proof.} We can prove by a straightforward computation that the
equation (\ref{eqEuLag1}) admits a standard Lagrangian description with 
\begin{equation*}
a(t,x)=\frac{{{{{\frac{\partial ^{2}\bar{\omega}}{\partial x^{2}}}}}}}{2{{{{%
\frac{\partial \bar{\omega}}{\partial x}}}}}},b(t,x)=\frac{{{\frac{\partial
^{2}\bar{\omega}}{\partial x\partial t}}}}{{{{{\frac{\partial \bar{\omega}}{%
\partial x}}}}}},c(t,x)=\frac{{{{{\frac{\partial \omega _{0}}{\partial x}-%
\frac{\partial \omega }{\partial t}+\frac{\partial ^{2}\bar{\omega}}{%
\partial t^{2}}}}}}}{2{{{{\frac{\partial \bar{\omega}}{\partial x}}}}}}.\Box
\end{equation*}

A top tangent form $\bar{\alpha}$ and a first order semi-spray $\bar{S}%
:I\!\!R\times TM\rightarrow I\!\!R\times T^{2}M$, having the local
expressions $\bar{\alpha}=\bar{\alpha}_{i}(t,x^{j},y^{j})dx^{i}$ and $\bar{S}%
(t,x^{j},y^{j})=(t,x^{i},y^{i},\bar{S}^{i}(t,x^{j},y^{j}))$ respectively,
give rise to a second order Lagrangian $L$, affine in accelerations, by the
formula 
\begin{equation*}
L(t,x^{i},y^{i},z^{i})=\bar{\alpha}_{i}\left( z^{i}-\bar{S}^{i}\right) .
\end{equation*}%
Let us suppose that there is a map $u:I\!\!R\times TM\rightarrow
I\!\!R\times TM$ of fibered manifolds over $I\!\!R\times M$, having the form 
$u(t,x^{i},y^{i})=(t,x^{i},u^{i}(t,x^{i},y^{i}))$, such that the semi-spray $%
\bar{S}$ has the local expression $\bar{S}%
^{i}(t,x^{i},y^{i})=u_{j}^{i}(t,x^{i},y^{i})y^{j}+u^{i}(t,x^{i},y^{i})$.
Then we can consider the tangent form $\omega \ $given by the formula. 
\begin{equation*}
\omega =\bar{\alpha}_{i}dy^{i}-\bar{\alpha}_{j}u_{i}^{j}dx^{i}-\bar{\alpha}%
_{j}u^{j}.
\end{equation*}

For example, a $2$--form $\alpha \in \mathcal{X}^{\ast }(M)\wedge \mathcal{X}%
^{\ast }(M)$, having the local expression $\alpha =\frac{1}{2}\alpha
_{ij}(x^{k})dx^{i}\wedge dx^{j}$, gives rise to a top tangent form $\bar{%
\alpha}=\alpha _{ij}y^{j}dx^{i}$. Adding a supplementary structure, one can
consider a tangent form. For example, if $\nabla $ is a linear connection on 
$M$, then one can $\bar{S}$ the spray associated with $\nabla $. Using local
coordinates, if $\{\Gamma _{jk}^{i}\}$ are the local coefficients of $\nabla 
$, then $\bar{S}^{i}=\frac{1}{2}\Gamma _{jk}^{i}y^{j}y^{k}$ are the local
coefficients of the first order spray. Then 
\begin{equation*}
\omega =\alpha _{ij}y^{j}dy^{i}-\alpha _{rj}y^{j}y^{k}\Gamma _{ik}^{r}dx^{i}.
\end{equation*}

A Riemannian metric $g$ on $M$ gives rise to the Levi-Civita connection $%
\nabla $. Using local coordinates, if $g=$ $\frac{1}{2}g_{ij}(x^{k})dx^{i}%
\otimes dx^{j}$, then $\Gamma _{jk}^{i}=g^{il}\Gamma _{ijk}$, where $(\Gamma
_{ijk})$ are the first order Christoffel coefficients $\Gamma _{klj}=$ $%
\frac{1}{2}\left( \frac{\partial g_{kj}}{\partial x^{l}}+\frac{\partial
g_{lj}}{\partial x^{k}}-\frac{\partial g_{kl}}{\partial x^{j}}\right) $ and $%
(g^{ij})=(g_{ij})^{-1}$. Then $\bar{S}^{i}(t,x^{i},y^{i})=g^{ij}\Gamma
_{klj}y^{k}y^{l}=u_{j}^{i}y^{j}$, $u_{j}^{i}=g^{ik}\Gamma _{ilk}y^{l}$.

The symplectic analogous version can be considered on a \emph{Fedosov
manifold}, i.e. a triple $(M,\alpha ,\nabla )$, where $(M,\alpha )$ is a
symplectic manifold and $\nabla $ is a symplectic linear connection on $M$,
i.e. $\alpha $ is parallel according the $\nabla $.

Let us consider the canonical symplectic form on $I\!\!R^{2r}$, $\alpha
^{(r)}=\varepsilon _{i,i+n}e^{i}\wedge e^{i+n}$, where $\left( \varepsilon
_{ij}\right) =\left( 
\begin{array}{cc}
0 & I_{r} \\ 
-I_{r} & 0%
\end{array}%
\right) $ is the Levi-Civita tensor on $I\!\!R^{2r}$. Then the second order
Lagrangian, affine in accelerations given by $L^{(2)}=\varepsilon
_{ij}y^{j}z^{i}+k\left\Vert y\right\Vert ^{2}$, where $\left\Vert
y\right\Vert ^{2}=\frac{1}{2}\delta _{ij}y^{i}y^{j}$. This Lagrangian
corresponds to some equivalent tangent forms $\omega =$ $\varepsilon
_{ij}y^{j}dy^{i}+k\delta _{ij}y^{j}dx^{i}$ and $\omega ^{\prime
}=\varepsilon _{ij}y^{j}dy^{i}+k\left\Vert y\right\Vert ^{2}$ (according to
Proposition \ref{preqg}). The tangent form $\omega $ is obtained using the
symplectic form $\left( \varepsilon _{ij}\right) $ and the semi-spray $\bar{S%
}$ on $I\!\!R^{2r}$ having the form $(t,x^{i},y^{i})\overset{\bar{S}}{%
\rightarrow }{}$ $(t,x^{i},y^{i},$ $\bar{S}^{i}=-y^{k}\delta
_{kj}\varepsilon ^{ji})$, where $(\varepsilon ^{ij})=(\varepsilon
_{ij})^{-1} $.

The second order Lagrangian, affine in accelerations given by $%
L^{(2)}=\varepsilon _{ij}y^{j}z^{i}+k\left\Vert y\right\Vert ^{2}$, where $%
\left\Vert y\right\Vert ^{2}=\frac{1}{2}\delta _{ij}y^{i}y^{j}$. This
Lagrangian corresponds to some equivalent tangent forms $\omega =$ $%
\varepsilon _{ij}y^{j}dy^{i}+k\delta _{ij}y^{j}dx^{i}$ and $\omega ^{\prime
}=\varepsilon _{ij}y^{j}dy^{i}+k\left\Vert y\right\Vert ^{2}$ (according to
Proposition \ref{preqg}). The tangent form $\omega $ is obtained using the
symplectic form $\left( \varepsilon _{ij}\right) $ and the semi-spray $\bar{S%
}$ on $I\!\!R^{2r}$ having the form $(t,x^{i},y^{i})\overset{\bar{S}}{%
\rightarrow }{}$ $(t,x^{i},y^{i},$ $\bar{S}^{i}=-ky^{k}\delta
_{kj}\varepsilon ^{ji})$, where $(\varepsilon ^{ij})=(\varepsilon
_{ij})^{-1}=-(\varepsilon _{ij})$.

\subsection{Tangent forms and first order semi-sprays}

We show below that in some special cases, the solutions of the generalized
Euler-Lagrange equation of a tangent form can be given by the integral
curves of certain local first order semi-sprays.

\textbf{Example 3.} Let us consider coordinates $(x,y)$ on $I\!\!R^{2}$ and $%
(x,y,X,Y)$ on $I\!\!R^{4}=TI\!\!R^{2}$. Let $\omega =(X+Y)dX+YdY$. As in
Example 1, the equations (\ref{EL-ord-2+}) have the solutions $x^{\prime
\prime \prime }(t)=y^{\prime \prime \prime }(t)=0$. Using the notations $%
x=x^{1}$, $y=x^{2}$, $X=y^{1}$, $Y=y^{2}$, then $\omega
=(y^{1}+y^{2})dy^{1}+y^{2}dy^{2}=$ $\bar{\omega}_{1}dy^{1}+\bar{\omega}%
_{2}dy^{2}$, $\left( h_{ij}\right) =\left( 
\begin{array}{cc}
0 & 1 \\ 
-1 & 0%
\end{array}%
\right) $ and $\left( h^{ij}\right) =\left( 
\begin{array}{cc}
0 & -1 \\ 
1 & 0%
\end{array}%
\right) $. The integral solutions of the vector field $X$ are $\frac{dx^{i}}{%
dt}=y^{i}$, $\frac{dy^{i}}{dt}=S^{i}=h^{ij}p_{j}$, $\frac{dp_{i}}{dt}=\tilde{%
\Phi}_{i}=0$. It follows that $p_{i}(t)=p_{i}^{0}$, thus $\frac{dx^{i}}{dt}%
=y^{i}$, $\frac{dy^{i}}{dt}=c_{i}$, where $c_{1}=-p_{i}^{0}$ and $%
c_{2}=p_{2}^{0}$. Finally we obtain all the solutions $\frac{d^{3}x^{i}}{%
dt^{2}}=0$. In conclusion, considering arbitrary semi-sprays on $I\!\!R^{2}$%
, with constant coefficients, then we obtain all the solutions of (\ref%
{EL-ord-2+}) as integral solutions of these first order semi-sprays.

The above example can be extended as follows.

\begin{pr}
\label{prex01}Let us suppose that there are some coordinates such that the
local coefficients of a regular tangent form $\omega $ depend only on $%
(y^{i})$. Then there is a family of local semi-sprays of first order whose
local coefficients depend only on $(y^{i})$, such that their integral curves
project on all the integral curves of the generalized Euler-Lagrange
equation of $\omega $.
\end{pr}

\emph{Proof.} The integral solutions of the vector field $X$ are $\frac{%
dx^{i}}{dt}=y^{i}$, $\frac{dy^{i}}{dt}=S^{i}=h^{ij}\left( p_{j}-\frac{%
\partial \omega _{0}}{\partial y^{i}}-\frac{\partial \omega _{j}}{\partial
y^{i}}y^{j}\right) $, $\frac{dp_{i}}{dt}=\tilde{\Phi}_{i}=-\frac{\partial
\omega _{i}}{\partial y^{j}}S^{j}$. Using the second equation in the
expression of the third, it follows that $\frac{dp_{i}}{dt}=-\frac{\partial
\omega _{i}}{\partial y^{j}}(y^{k})\frac{dy^{j}}{dt}$, thus $p_{i}+\omega
_{i}=c_{i}$ along every solution. It follows that if considering local
semi-sprays having as local component functions $\bar{S}%
^{i}(y^{k})=h^{ij}(y^{k})\left( c_{j}-\omega _{j}(y^{k})-\frac{\partial
\omega _{0}}{\partial y^{i}}(y^{k})-\frac{\partial \omega _{j}}{\partial
y^{i}}(y^{k})y^{j}\right) $, we obtain all the integral curves of the
generalized Euler-Lagrange equation of $\omega $. $\Box $

Since the tangent form $\omega ^{\prime }=\omega +dF$ has the same extrema
curves as $\omega $, the extrema curves of the tangent forms $\omega
^{\prime }=\left( \omega _{0}(y^{j})+\frac{\partial F}{\partial t}\right)
dt+\left( \omega _{i}(y^{j})+\frac{\partial F}{\partial x^{i}}\right)
dx^{i}+\left( \bar{\omega}_{i}(y^{j})+\frac{\partial F}{\partial y^{i}}%
\right) dy^{i}$ and $\omega =\omega _{0}(y^{j})dt+\omega _{i}(y^{j})dx^{i}+%
\bar{\omega}_{i}(y^{j})dy^{i}$ (used in Proposition above) are the same. In
order to detect when one can apply the Proposition above, we prove the
following result.

\begin{pr}
\label{prex02}Let us consider a tangent form $\mu $, a point $x_{0}\in M$
and a local system of coordinates $(U,\varphi )$, where $x_{0}\in U$. Then
the following statements are equivalent:

\begin{enumerate}
\item There is a local tangent form $\omega =\mu -dF$ on a $TU^{\prime }$, $%
x_{0}\in U^{\prime }\subset U$, such that the local components of $\omega $
does depend only on $(y^{i})$.

\item The local components of $d\mu $ depend only on $(y^{i})$ and the
components of $\{dx^{i}\wedge dt$, $dx^{i}\wedge dx^{j}\}$ vanish.
\end{enumerate}
\end{pr}

\emph{Proof.} If the property 1. holds for $\mu $, then $d\mu =d\omega $,
thus 2. follows. Conversely, let us suppose that 2. holds, thus $d\mu
=a_{i}(y^{k})dy^{i}\wedge dt+$ $b_{ij}(y^{k})dx^{i}\wedge dy^{j}+$ $\tfrac{1%
}{2}c_{ij}(y^{k})dy^{i}\wedge dy^{j}$. Then we have $0=dd\mu =$ $\frac{%
\partial a_{i}}{\partial y^{k}}dy^{k}\wedge dy^{i}\wedge dt+$ $\frac{%
\partial b_{ij}}{\partial y^{k}}dy^{k}\wedge dx^{i}\wedge dy^{j}+$ $\tfrac{1%
}{2}\frac{\partial c_{ij}}{\partial y^{k}}dy^{k}\wedge dy^{i}\wedge dy^{j}$.
Thus using the Poincar\'{e} Lemma, it follows that $a_{i}=\frac{\partial f}{%
\partial y^{i}}$, $b_{ij}=\frac{\partial g_{i}}{\partial y^{j}}$ and $c_{ij}=%
\frac{\partial h_{i}}{\partial y^{j}}-\frac{\partial h_{j}}{\partial y^{i}}$
on $I\!\!R^{m}$, where $f$, $g_{i}$, $h_{i}:I\!\!R^{m}\rightarrow I\!\!R$
are functions that depend only on $(y^{i})$. Let us consider the form $%
\omega =fdt+g_{i}dx^{i}+h_{i}dy^{i}$ on $TU=U\times I\!\!R^{m}$. Then $d\mu
=d\omega $, or $d(\mu -\omega )=0$, thus for a sufficiently small $U^{\prime
}\subset U$, $x_{0}\in U^{\prime }$, one have $\mu -\omega =dF$ on $%
TU^{\prime }$. $\Box $

\textbf{Example 4}. Consider the tangent form $\omega =-ydx+xdy+YdX-XdY$ on $%
I\!\!R^{2}$ used in Example 2., with coordinates $(x,y)$ on $I\!\!R^{2}$ and 
$(x,y,X,Y)$ on $I\!\!R^{4}=TI\!\!R^{2}$. We use below also the notations $%
x=x^{1}$, $y=x^{2}$, $X=y^{1}$, $Y=y^{2}$, then $\omega
=-x^{2}dx^{1}+x^{1}dx^{2}+y^{2}dy^{1}-y^{1}dy^{2}=$ $\omega
_{1}dx^{1}+\omega _{2}dx^{2}+\bar{\omega}_{1}dy^{1}+\bar{\omega}_{2}dy^{2}$, 
$\left( h_{ij}\right) =\left( 
\begin{array}{cc}
0 & 2 \\ 
-2 & 0%
\end{array}%
\right) $ and $\left( h^{ij}\right) =\left( 
\begin{array}{cc}
0 & -1/2 \\ 
1/2 & 0%
\end{array}%
\right) $. The integral solutions of the vector field $X$ are $\frac{dx^{i}}{%
dt}=y^{i}$, $\frac{dy^{i}}{dt}=S^{i}=h^{ij}p_{j}$, $\frac{dp_{i}}{dt}=\tilde{%
\Phi}_{i}=\left( \frac{\partial \omega _{j}}{\partial x^{i}}-\frac{\partial
\omega _{i}}{\partial x^{j}}\right) {y}^{j}$. Specifically, $\frac{dp_{1}}{dt%
}=2y^{2}=2\frac{dx^{2}}{dt}$ and $\frac{dp_{2}}{dt}=-2y^{1}=-2\frac{dx^{1}}{%
dt}$. Thus $p_{1}=2x^{2}+2c_{1}$ and $p_{2}=$ $-2x^{1}+2c_{2}$. Considering
the local first order semi-sprays $\bar{S}^{1}(x^{i},y^{i})=x^{2}+c_{1}$ and 
$\bar{S}^{2}(x^{i},y^{i})=-x^{1}+c_{2}$, we obtain the system $\frac{dx^{i}}{%
dt}=y^{i}$, $\frac{dy^{i}}{dt}=\bar{S}^{i}$.. Taking into account the
Example 2., the integral curves of all semi-sprays $\bar{S}$ having this
form give all the solutions of the generalized Euler-Lagrange equation (\ref%
{EL-ord-2+}) of $\omega $.

The above example can be extended as follows.

\begin{pr}
\label{prex03}Let us suppose that there are some coordinates such that the
local expression of a regular tangent form $\omega $ is $\omega =\omega
_{0}(y^{j})+\omega _{i}(x^{j})dx^{i}+\bar{\omega}_{i}(y^{j})dy^{i}$ and $%
\frac{\partial \omega _{i}}{\partial x^{j}}+\frac{\partial \omega _{j}}{%
\partial x^{i}}=0$. Then there is a family of local semi-sprays of first
order, such that their integral curves project on all the integral curves of
the generalized Euler-Lagrange equation of $\omega $.
\end{pr}

\emph{Proof.} The integral solutions of the vector field $X$ are $\frac{%
dx^{i}}{dt}=y^{i}$, $\frac{dy^{i}}{dt}=S^{i}=h^{ij}\left( p_{j}-\frac{%
\partial \omega _{0}}{\partial y^{i}}\right) $, $\frac{dp_{i}}{dt}=\tilde{%
\Phi}_{i}=\left( \frac{\partial \omega _{j}}{\partial x^{i}}-\frac{\partial
\omega _{i}}{\partial x^{j}}\right) y^{j}=$ $-2\frac{\partial \omega _{i}}{%
\partial x^{j}}y^{j}$. Using Lemma \ref{lmhelp} below, $\{\omega _{i}\}$
have the form $\omega _{i}=c_{ij}x^{j}+d_{i}$, thus $%
p_{i}=-2c_{ij}x^{j}+e_{j}$, where $c_{ij}=-c_{ji}$, $d_{i}$ and $e_{i}$ are
constants. It follows that considering local semi-sprays having as local
component functions $\bar{S}^{i}(y^{k})=h^{ij}(y^{k})\left(
-2c_{ij}x^{j}+e_{j}-\frac{\partial \omega _{0}}{\partial y^{i}}%
(y^{k})\right) $, for all constants $\{e_{j}\}$, we obtain all the integral
curves of the generalized Euler-Lagrange equation of $\omega $. $\Box $

\begin{lm}
\label{lmhelp}Given the set $\{\omega _{i}(x^{j})\}_{i=\overline{1,m}}$ of
real functions on $I\!\!R^{m}$, then the following conditions are equivalent:

\begin{enumerate}
\item $\frac{\partial \omega _{i}}{\partial x^{j}}+\frac{\partial \omega _{j}%
}{\partial x^{i}}=0$, $i,j=\overline{1,m}$;

\item there are constants $\{c_{ij},d_{i}\}_{i,j=\overline{1,m}}$, $%
c_{ij}=-c_{ji}$ such that $\omega _{i}=c_{ij}x^{j}+d_{i}$;

\item there is a set $\{\varphi _{i}\}_{i=\overline{1,m}}$ of real functions
on $I\!\!R^{m}$ such that $\frac{\partial \omega _{i}}{\partial x^{j}}-\frac{%
\partial \omega _{j}}{\partial x^{i}}=\frac{\partial \varphi _{i}}{\partial
x^{j}}$.
\end{enumerate}
\end{lm}

\emph{Proof.} Obviously 2. implies 1. and 3. Let us suppose that 1. holds.
We have $\omega _{ij}=\frac{\partial \omega _{i}}{\partial x^{j}}-\frac{%
\partial \omega _{j}}{\partial x^{i}}=$ $2\frac{\partial \omega _{i}}{%
\partial x^{j}}$; then $\frac{\partial \omega _{ij}}{\partial x^{k}}=2\frac{%
\partial ^{2}\omega _{i}}{\partial x^{j}\partial x^{k}}=\frac{\partial
\omega _{ik}}{\partial x^{j}}$, thus $\frac{\partial ^{2}\omega _{i}}{%
\partial x^{j}\partial x^{k}}-\frac{\partial ^{2}\omega _{j}}{\partial
x^{i}\partial x^{k}}=$ $\frac{\partial ^{2}\omega _{i}}{\partial
x^{k}\partial x^{j}}-\frac{\partial ^{2}\omega _{k}}{\partial x^{i}\partial
x^{j}}$, that gives $\frac{\partial \omega _{jk}}{\partial x^{i}}=0$, thus
2. holds. Let us suppose that 3. holds. We have $\frac{\partial ^{2}\omega
_{i}}{\partial x^{j}\partial x^{k}}-\frac{\partial ^{2}\omega _{j}}{\partial
x^{i}\partial x^{k}}=$ $\frac{\partial ^{2}\omega _{i}}{\partial
x^{k}\partial x^{j}}-\frac{\partial ^{2}\omega _{k}}{\partial x^{i}\partial
x^{j}}$, thus 3. holds as previously. $\Box $

The tangent form $\omega ^{\prime }=\omega +dF$ has the same extrema curves
as $\omega $. Thus the extrema curves of the tangent forms $\omega ^{\prime
}=\frac{\partial F}{\partial t}dt+\left( \omega _{i}(x^{j})+\frac{\partial F%
}{\partial x^{i}}\right) dx^{i}+\left( \bar{\omega}_{i}(y^{j})+\frac{%
\partial F}{\partial y^{i}}\right) dy^{i}$ and $\omega $ from Proposition
above are the same. In particular, one can relax the hypothesis of
Proposition above, asking the existence of a local function $F(x^{i})$ such
that $\frac{\partial \omega _{i}}{\partial x^{j}}+\frac{\partial \omega _{j}%
}{\partial x^{i}}=-2\frac{\partial ^{2}F}{\partial x^{i}\partial x^{j}}$;
more precisely, $\omega _{i}+\frac{\partial F}{\partial x^{i}}%
=c_{ij}x^{j}+d_{i}$, where $c_{ij}=-c_{ij}\ $and $d_{i}$ are constants. In
order to apply the result from Proposition above, we prove the following
result.

\begin{pr}
\label{prex04}Let us consider a tangent form $\mu $, a point $x_{0}\in M$
and a local system of coordinates $(U,\varphi )$, where $x_{0}\in U$. Then
the following statements are equivalent:

\begin{enumerate}
\item There is a local tangent form $\omega =\mu -dF$ on a $TU^{\prime }$, $%
x_{0}\in U^{\prime }\subset U$, such that $\omega =\omega _{i}(x^{j})dx^{i}+%
\bar{\omega}_{i}(y^{j})dy^{i}$ and $\frac{\partial \omega _{i}}{\partial
x^{j}}+\frac{\partial \omega _{j}}{\partial x^{i}}=0$.

\item The local components of $d\mu $ have the properties that the
components of $\{dx^{i}\wedge dt$, $dy^{i}\wedge dt$, $dx^{i}\wedge dy^{j}\}$
vanish, the components of $\{dy^{i}\wedge dy^{j}\}$ depend only on $(y^{i})$
and the components $\mu _{ij}$ of $\{dx^{i}\wedge dx^{j}\}$ are constants.
\end{enumerate}
\end{pr}

\emph{Proof.} If the property 1. holds for $\mu $, then $d\mu =d\omega $,
thus 2. follows. Conversely, let us suppose that 2. holds, thus $d\mu =%
\tfrac{1}{2}\mu _{ij}dx^{i}\wedge dx^{j}+$ $\tfrac{1}{2}\nu
_{ij}(y^{k})dy^{i}\wedge dy^{j}$, with $\mu _{ij}=-\mu _{ji}\ $constants.
Using $dd\mu =0$ and the Poincar\'{e} Lemma, it follows that $\nu _{ij}=%
\frac{\partial g_{i}}{\partial y^{j}}(y^{k})-\frac{\partial g_{j}}{\partial
y^{i}}(y^{k})$ , where $g_{i}:I\!\!R^{m}\rightarrow I\!\!R$. Let us denote $%
f_{i}(x^{k})=\mu _{ij}x^{j}$ and consider the local differential form $%
\omega =f_{i}(x^{k})dx^{i}+g_{i}(y^{k})dy^{i}$ on $TU=U\times I\!\!R^{m}$.
Then $d\mu =d\omega $, or $d(\mu -\omega )=0$, thus for a sufficiently small 
$U^{\prime }\subset U$, $x_{0}\in U^{\prime }$, one have $\mu -\omega =dF$
on $TU^{\prime }$. $\Box $

The tangent forms $\omega =\omega _{i}(x^{j})dx^{i}+\bar{\omega}%
_{i}(y^{j})dy^{i}$ and $\omega ^{\prime }=y^{i}\omega _{i}(x^{j})dt+\bar{%
\omega}_{i}(y^{j})dy^{i}$ are equivalent. If $\omega _{i}(x^{j})=c_{ij}x^{j}$%
, then $d\omega ^{\prime }=c_{ij}x^{j}dy^{i}\wedge dt+$ $c_{ij}y^{i}dx^{j}%
\wedge dt+$ $\frac{\partial \bar{\omega}_{i}}{\partial y^{j}}%
(y^{j})dy^{i}\wedge dy^{j}$. Then the following result follows in the same
line as the previous ones.

\begin{pr}
\label{prex05}Let us consider a tangent form $\mu $, a point $x_{0}\in M$
and a local system of coordinates $(U,\varphi )$, where $x_{0}\in U$. Then
the following statements are equivalent:

\begin{enumerate}
\item There is a local tangent form $\omega =\mu -dF$ on a $TU^{\prime }$, $%
x_{0}\in U^{\prime }\subset U$, such that $\omega =\omega _{i}(x^{j})y^{i}dt+%
\bar{\omega}_{i}(y^{j})dy^{i}$ and $\frac{\partial \omega _{i}}{\partial
x^{j}}+\frac{\partial \omega _{j}}{\partial x^{i}}=0$.

\item The local components of $d\mu $ have the properties that the
components of $\{dx^{i}\wedge dx^{j}$, $dx^{i}\wedge dy^{j}\}$ vanish, the
components of $\{dy^{i}\wedge dy^{j}\}$ depend only on $(y^{i})$ and the
components $f_{i}$ of $\{dx^{i}\wedge dt\}$ and $g_{j}$ of $\{dy^{i}\wedge
dt\}$have the property that $\frac{\partial f_{j}}{\partial y^{i}}=\frac{%
\partial g_{i}}{\partial h^{j}}=c_{ij}$ are constants.
\end{enumerate}
\end{pr}

\section{Appendix}

As a manifold, $T^{2}M\subset TTM$ is the submanifold of the vectors $X_{v}$
that project according to the double vector bundle structure $\pi
^{(2)}:TTM\rightarrow TM$, as tangent bundle of $TM$ and $\pi _{\ast
}^{(1)}:TTM\rightarrow TM$, as the differential of the canonical projection $%
\pi ^{(1)}:TM\rightarrow M$). As a manifold, the point in $T^{2}M$ can be
defined also as the equivalent classes of curves on $M$ having a $2$%
--contact in a point (see, for example \cite{Kra, Le, Mi}).

A \emph{slashed} (first order) Lagrangian on $M$ is a differentiable map $%
L:TM_{\ast }\rightarrow I\!\!R$, where $TM_{\ast }=TM\backslash \{0\}$ and $%
\{0\}$ is the image of the null section $M\rightarrow TM$. Analogously, a
slashed second order Lagrangian on $M$ is a differentiable map $%
L^{(2)}:T^{2}M_{\ast }\rightarrow I\!\!R$, where $T^{2}M_{\ast
}=T^{2}M\backslash \{0\}$ and $\{0\}$ is the image of the
,,null\textquotedblright\ section $M\rightarrow T^{2}M$ given by $%
(x^{i})\rightarrow (x^{i}$, $y^{i}=0$, $z^{i}=0)$.

Coordinates $(x^{i})$ on $M$, $(x^{i},y^{i})$ on $TM$, $%
(x^{i},y^{i},X^{i},Y^{i})$ on $TM$ and $(x^{i},y^{i},z^{i})$ on $T^{2}M$
follow the rules $x^{i^{\prime }}=x^{i^{\prime }}(x^{i})$, $y^{i^{\prime }}=%
\frac{\partial x^{i^{\prime }}}{\partial x^{i}}y^{i}$, $X^{i^{\prime }}=%
\frac{\partial x^{i^{\prime }}}{\partial x^{i}}X^{i}$, $Y^{i^{\prime }}=y^{j}%
\frac{\partial ^{2}x^{i^{\prime }}}{\partial x^{j}\partial x^{i}}X^{i}+\frac{%
\partial x^{i^{\prime }}}{\partial x^{i}}Y^{i}$, $z^{i^{\prime }}={{\frac{1}{%
2}}}y^{i}y^{j}\frac{\partial ^{2}x^{i^{\prime }}}{\partial x^{j}\partial
x^{i}}+\frac{\partial x^{i^{\prime }}}{\partial x^{i}}z^{i}$ on an
intersection domain (see, for example, \cite{Mi}). It follows that some
local coordinates $(x^{i},p_{i})$ on $T^{\ast }M$ and $%
(x^{i},y^{i},P_{i},p_{i})$ on $T^{\ast }TM$ follow the next rules on a
common domain of coordinates: $x^{i^{\prime }}$ and $y^{i^{\prime }}$ as
above, $p_{i}=p_{i^{\prime }}\frac{\partial x^{i^{\prime }}}{\partial x^{i}}$
and $P_{i}=\frac{\partial y^{i^{\prime }}}{\partial x^{i}}p_{i^{\prime }}+%
\frac{\partial x^{i^{\prime }}}{\partial x^{i}}P_{i}$. There is a natural
flip $\iota :TT^{\ast }M\rightarrow T^{\ast }TM$ that is a diffeomorphism;
it has the local expression $(x^{i},y^{i},P_{i},p_{i})\overset{\iota }{%
\rightarrow }{}(x^{i},p_{i},y^{i},P_{i})$.

On $T^{2}M$ it can be also considered the coordinates $%
(x^{i},y^{(1)i}=y^{i},y^{(2)i}=z^{i})$, as well as $(x^{i},\dot{x}^{i},\ddot{%
x}^{i})$, that are more suitable for expressing the derivatives of the
functions. The connections between the coordinates $(x^{i},\dot{x}^{i},\ddot{%
x}^{i})$ and $(x^{i},y^{i},z^{i})$ are $x^{i}=x^{i}$, $\dot{x}^{i}=y^{i}$,
but $\ddot{x}^{i}=2z^{i}$. Notice that using local coordinates, the
inclusion $T^{2}M\subset TTM$ has the expression $(x^{i},y^{i},z^{i})%
\rightarrow $ $(x^{i},y^{i},y^{i},z^{i})$.

There are affine bundles structures $\pi _{1}^{2}:T^{2}M\rightarrow TM$ and $%
\pi _{2}^{3}:T^{3}M\rightarrow T^{2}M$; in general $\pi
_{k}^{k+1}:T^{k}M\rightarrow T^{k-1}M$, $k\geq 2$. A (time independent)
semi-spray of order $k$ is a section $S:T^{k}M\rightarrow T^{k+1}M$.
Considering the product bundle $I\!\!R\times T^{k+1}M\rightarrow
I\!\!R\times T^{k}M$, $k\geq 1$, then a (time dependent) semi-spray of order 
$k$ is a section $S:I\!\!R\times T^{k}M\rightarrow I\!\!R\times T^{k+1}M$,
such that $S(t,\bar{x})=(t,\bar{x},(k+1)S^{i}(t,\bar{x}))$; this semi-spray
of order $k$ is considered in the paper.

The integral curves of a $k$-order semi-spray $S$ are exactly the integral
curves of $S$ regarded as a vector field on $T^{k}M$. Using coordinates $%
(x^{i},y^{(1)i},\ldots ,y^{(k)i})$ on $T^{k}M$, the local expression of a $k$%
-order semi-spray is $S=y^{(1)i}{\ {\frac{\partial }{\partial x^{i}}}}%
+2y^{(2)i}{\ {\frac{\partial }{\partial y^{(1)i}}}}+\cdots +ky^{(k)i}{\ {%
\frac{\partial }{\partial y^{(k-1)i}}}}-(k+1)S^{i}(x^{i},y^{(1)i},\ldots
,y^{(k)i})$. We say that $S^{i}$ are the local functions that give $S$.

Consider now a tangent form $\omega \in \mathcal{X}^{\ast }(I\!\!R\times TM)$
given in local coordinates by (\ref{eqlocom}). Then $\omega
_{0}:I\!\!R\times TM\rightarrow I\!\!R$ gives a (global defined) real
function. According to two couples of coordinates $(x^{j},y^{j})$ and $%
(x^{j^{\prime }},y^{j^{\prime }})$ on the common domain, the local
components $\omega _{i}$ and $\bar{\omega}_{i}$ follow the rules $\bar{\omega%
}_{i}={{\frac{\partial x^{i^{\prime }}}{\partial x^{i}}}}\bar{\omega}%
_{i^{\prime }}$ and $\omega _{i}={{\frac{\partial y^{i^{\prime }}}{\partial
x^{i}}}}\bar{\omega}_{i^{\prime }}+{{\frac{\partial x^{i^{\prime }}}{%
\partial x^{i}}}}\omega _{i^{\prime }}$ respectively. We can consider the
top components $(\bar{\omega}_{i})$ defining a section $\bar{\omega}%
:I\!\!R\times TM\rightarrow \pi ^{\ast }T^{\ast }M$, $\bar{\omega}=\bar{%
\omega}_{i}(t,x^{j},y^{j})dx^{i}$, of the induced vector bundle $\pi
_{1}=\pi ^{\ast }(\pi ^{\prime }):\pi ^{\ast }T^{\ast }M\rightarrow
I\!\!R\times TM$, where $\pi :I\!\!R\times TM\rightarrow M$ comes from the
tangent bundle $TM\rightarrow M$ and $\pi ^{\prime }:T^{\ast }M\rightarrow M$
is the cotangent bundle of $M$. In general, a section $\bar{\omega}%
:I\!\!R\times TM\rightarrow \pi ^{\ast }T^{\ast }M$ is a top tangent form
(on $M$). 

We say that a Lagrangian $L:I\!\!R\times TM\rightarrow I\!\!R$ is \emph{%
pointed} if $L(t,x^{i},y^{i}=0)=0$.

\begin{pr}
A Lagrangian $L:I\!\!R\times TM\rightarrow I\!\!R$ is a pointed one iff
there is to top tangent form $\nu =\nu _{i}(t,x^{i},y^{i})dx^{i}$, such that 
$L(t,x^{i},y^{i})=y^{i}\nu _{i}$.
\end{pr}

\emph{Proof.} The sufficiency is obvious, we prove only the necessity.
Indeed, if $L$ is pointed, then $L(t,x^{i},y^{i})=y^{i}\int_{0}^{1}\frac{%
\partial L}{\partial y^{i}}(t,x^{i},\tau y^{i})d\tau =y^{i}\nu _{i}$. It can
be easily checked that $\nu =\nu _{i}dx^{i}$ is a global top tangent form. $%
\Box $

An other example of a top tangent form: if $L^{(2)}:T^{2}M\rightarrow I\!\!R$
is a second order Lagrangian, affine in accelerations, then $\bar{\omega}=%
\frac{\partial L^{(2)}}{\partial z^{i}}dx^{i}$ is a top tangent form. Notice
that a top tangent form $\bar{\omega}=\bar{\omega}_{i}(t,x^{j},y^{j})dx^{i}$
is a degenerated tangent form. Since $\bar{\omega}=\bar{\omega}_{i}dt+\bar{%
\omega}_{i}(dx^{i}-y^{i}dt)$, it follows that $\bar{\omega}$ is equivalent
to the first order (pointed) Lagrangian $L_{0}=\bar{\omega}_{i}y^{i}$.
Conversely, it is easy to see that a pointed Lagrangian $L_{0}=\bar{\omega}%
_{i}(t,x^{j},y^{j})y^{i}$ is equivalent to the top tangent form $\bar{\omega}%
=\bar{\omega}_{i}dx^{i}$.

An analogous object considered in the paper is a \emph{pure tangent form}
that can be considered as a section $\omega ^{\prime }:I\!\!R\times
TM\rightarrow \pi _{0}^{\ast }T^{\ast }TM$, $\omega ^{\prime }=\bar{\omega}%
_{i}(t,x^{j},y^{j})dy^{i}+\omega _{i}(t,x^{j},y^{j})dx^{i}$, of the induced
vector bundle $\pi _{2}=\pi _{0}^{\ast }(\pi ^{\prime \prime }):\pi
_{0}^{\ast }T^{\ast }TM\rightarrow I\!\!R\times TM$, where $\pi ^{\prime
\prime }:T^{\ast }TM\rightarrow TM$ is the cotangent bundle of $TM$ and $\pi
_{0}:I\!\!R\times TM\rightarrow TM$ is the trivial projection.

A \emph{fibered manifold} is a surjective submersion $\pi _{E}:E\rightarrow M
$; $E_{x}=\pi _{M}^{-1}(x)$ is the \emph{fiber} of $x\in M$. There are local
coordinates, called as \emph{adapted} to the submersion (or to the fibered
manifold structure), giving the local form $(x^{i},u^{\alpha })\overset{\pi
_{E}}{\rightarrow }{}(x^{i})$. Considering coordinates $(x^{i},u^{\alpha })$
and $(x^{i^{\prime }},u^{\alpha ^{\prime }})$ on an intersection domain,
then $x^{i^{\prime }}=x^{i^{\prime }}(x^{i})$ and $u^{\alpha ^{\prime
}}=u^{\alpha ^{\prime }}(x^{i},u^{\alpha })$.

A \emph{fibered map} of two fibered manifolds $\pi _{E}:E\rightarrow M$ and $%
\pi _{F}:F\rightarrow M^{\prime }$ is a couple $(f_{0}:M\rightarrow
M^{\prime },f:E\rightarrow E^{\prime })$ of maps that send fibers to fibers,
i.e. $f_{0}\circ \pi _{E}=\pi _{F}\circ f$. If $M=M^{\prime }$ and $%
f_{0}=id_{M}$, then $f$ is called simply a fibered map (of fibered manifolds
over the same base). Let $\pi _{E}:E\rightarrow M$ and $\pi
_{F}:F\rightarrow M$ be fibered manifolds over the same base. The fibered
manifold product $P=E\times _{M}F$ is $P=\underset{x\in M}{\cup }%
{}P_{x}\subset E\times F$, where $P_{x}=\{(e,f)\in E\times F:$ $\pi
_{E}(e)=\pi _{F}(f)\}$ is a new fibered manifold $\pi _{P}:P\rightarrow M$
over $M$, but also over $E$ and $F$. The tangent space of $P$ is locally the
sum of two subspaces, each tangent to two foliations. Using coordinates, we
explicit in two cases, useful in the paper. First is when $E=TM$ and $%
F=T^{\ast }M$ are the tangent and the cotangent space of $M$ respectively.
In this case, considering $(x^{i})$, $(x^{i},y^{i})$, $(x^{i},p_{i})$, $%
(x^{i},y^{i},p_{i})$ and $(x^{i},y^{i},p_{i},X^{i},Yi,P_{i})$ local
coordinates on $M$, $TM$, $T^{\ast }M$, $TM\times T^{\ast }M$ and $%
T(TM\times T^{\ast }M)$ respectively, then these coordinates follow the next
rules on a common domain: $x^{i^{\prime }}=x^{i^{\prime }}(x^{i})$, $%
y^{i^{\prime }}=\frac{\partial x^{i^{\prime }}}{\partial x^{i}}y^{i}$, $%
p_{i}=\frac{\partial x^{i^{\prime }}}{\partial x^{i}}p_{i^{\prime }}$, $%
X^{i^{\prime }}=\frac{\partial x^{i^{\prime }}}{\partial x^{i}}X^{i}$, $%
Y^{i^{\prime }}=\frac{\partial y^{i^{\prime }}}{\partial x^{i}}X^{i}+\frac{%
\partial x^{i^{\prime }}}{\partial x^{i}}Y^{i}$ and $P_{i}=\frac{\partial
y^{i^{\prime }}}{\partial x^{i}}p_{i^{\prime }}+\frac{\partial x^{i^{\prime
}}}{\partial x^{i}}P_{i^{\prime }}$ respectively. A second case is when $%
E=F=T^{\ast }M$ and $T^{\ast }M\times _{M}T^{\ast }M=T_{2}^{0}M$;
considering some local coordinates $%
(x^{i},p_{(0)i},p_{(1)i},y^{i},P_{(0)i},P_{(1)i})$ on $TT_{2}^{0}M$, then $%
p_{(0)i}=\frac{\partial x^{i^{\prime }}}{\partial x^{i}}p_{(0)i^{\prime }}$, 
$p_{(0)i}=\frac{\partial x^{i^{\prime }}}{\partial x^{i}}p_{(0)i^{\prime }}$%
, $y^{i^{\prime }}=\frac{\partial x^{i^{\prime }}}{\partial x^{i}}y^{i}$, $%
P_{(0)i}=\frac{\partial y^{i^{\prime }}}{\partial x^{i}}p_{(0)i^{\prime }}+%
\frac{\partial x^{i^{\prime }}}{\partial x^{i}}P_{(0)i^{\prime }}$, $%
P_{(1)i}=\frac{\partial y^{i^{\prime }}}{\partial x^{i}}p_{(1)i^{\prime }}+%
\frac{\partial x^{i^{\prime }}}{\partial x^{i}}P_{(1)i^{\prime }}$.

If $\pi _{E}:E\rightarrow M$ is a fibered manifold and $f_{0}:M^{\prime
}\rightarrow M$ is a differentiable map, then $f_{0}^{\ast }E=\{(x^{\prime
},e)\in M^{\prime }\times E:f_{0}(x^{\prime })=\pi _{E}(e)\}$ is a
differentiable manifold. The canonical projections $\pi _{f_{0}^{\ast
}E}=\pi _{1}:f_{0}^{\ast }E\rightarrow M^{\prime }$ and $f=\pi
_{2}:f_{0}^{\ast }E\rightarrow E$ give a fibered manifold $(f_{0}^{\ast
}E,\pi _{f_{0}^{\ast }E},M^{\prime })$ and a fibered map $(f_{0},f)$.

Paul Popescu\\
University of Craiova\\
Department of Applied Mathematics\\
13, Al.I.Cuza st., Craiova, 200585, Romania\\
E-mail: Paul\_P\_Popescu@yahoo.com

\end{document}